%% file: jpsi_prc.tex
\documentclass[%
reprint,
superscriptaddress,
nofootinbib,
amsmath,amssymb,
prc,
floatfix
]{revtex4-1}

\usepackage{graphicx}
\usepackage{setspace}
\usepackage{dcolumn}
\usepackage{bm}
\usepackage{hyperref}
\usepackage{xcolor}
\usepackage[caption=false]{subfig}
\usepackage{orcidlink}

\begin{document}

\preprint{Draft}

\title{Measurement of the J/$\psi $ photoproduction cross section over
the full near-threshold kinematic region}

\input{authors.tex}

\date{\today}

\begin{abstract}

We report the total and differential cross sections for $J/\psi$ photoproduction with the large acceptance GlueX spectrometer for photon beam energies from the threshold at 8.2~GeV up to 11.44~GeV and over the full kinematic range of momentum transfer squared, $t$. 
Such coverage facilitates the extrapolation of the differential cross sections to the forward ($t = 0$) point beyond the physical region. The forward cross section is used by many theoretical models and plays an important role in understanding $J/\psi$ photoproduction and its relation to the $J/\psi-$proton interaction. 
These measurements of $J/\psi$ photoproduction near threshold are also crucial inputs to theoretical models that are used to study important aspects of the gluon structure of the proton, such as the gluon Generalized Parton Distribution (GPD) of the proton, the mass radius of the proton, and the trace anomaly contribution to the proton mass.  
We observe possible structures in the total cross section energy
dependence and find evidence for contributions beyond gluon exchange
in the differential cross section close to threshold, both of which
are consistent with contributions from open-charm intermediate states.

\end{abstract}

\maketitle

\section{Introduction}

Over the past several years there has been a renewed interest 
in studying near-threshold $J/\psi $ photoproduction as a tool to experimentally probe 
important properties of the nucleon related to its gluon content. 
Such experiments became possible thanks to the $12$~GeV upgrade
of the CEBAF accelerator at Jefferson Lab
covering the threshold region of the reaction,
resulting in the first exclusive measurements
very close to threshold by the GlueX collaboration~\cite{prl_gluex}.

Exclusive $J/\psi $ photoproduction is expected to proceed dominantly 
through gluon exchange due to the heavy mass of the charm quark.
Thus, the $t$ dependence of the reaction is defined by the proton vertex, 
which provides a probe of the nucleon gluon form factors~\cite{strikman}. 
The extraction of the gluonic properties of the proton from $J/\psi $ production data
requires additional assumptions. 
One such assumption is the use of Vector Meson Dominance (VMD) 
to relate the $\gamma p \to J/\psi p$ reaction 
to elastic $J/\psi p \to J/\psi p$ scattering. 
At low energies, the latter reaction is related to several fundamental quantities.
These include the trace anomaly contribution to the mass of the proton~\cite{Kharzeev99,Hatta1,Wang2}, 
and the $J/\psi p$ scattering length which is related to the 
possible existence of a charmonium-nucleon bound state~\cite{SL_1,SL_2}.

An important QCD approach is to assume factorization between the
gluon Generalized Parton Distributions (gGPD) of the proton and the $J/\psi $ 
wave function, and the hard quark-gluon interaction.
The hard scale in this approach is defined by the heavy quark mass.
In Ref.~\cite{Ivanov} such a general approach was applied to the $J/\psi $ photoproduction
in leading-order (LO) and next-to-leading-order (NLO) 
at high energy and small transferred momentum $|t|$. 
An important continuation of these efforts can be found in Refs.~\cite{Hatta_Strikman,Ji2021},
where it was shown in LO and for heavy quark masses, 
that factorization also holds at energies down to threshold
for large absolute values of $t$. 
Close to threshold, due to the large skewness parameter, the spin-2 
(graviton-like) two-gluon exchange 
dominates ~\cite{Hatta_Strikman} and therefore $J/\psi$ photoproduction can be used 
to study the gravitational form factors of the proton~\cite{Ji2021}. 
Such information was used to estimate the {\it mass}
radius of the proton~\cite{Kharzeev21,Jimass,Zahed2, Wang1}, 
as opposed to the well-known \textit{charge} radius.
Alternatively, the holographic approach was used to describe the soft part
of $J/\psi $ photoproduction and relate the differential cross sections
to the gravitational form factors~\cite{Hatta1, Hatta2, Zahed1, Zahed2, Zahed3}.

However, such an ambitious program to study the mass properties of the proton
requires detailed investigation of the
above assumptions used to interpret the data.
Ref.~\cite{Feng2} calculates directly Feynman diagrams
of the near threshold heavy quarkonium photoproduction
at large momentum transfer and finds that there is no direct connection
to the gravitational form factors.
In contrast to the above gluon-exchange mechanisms, it was proposed 
in Ref.~\cite{openc} that $J/\psi $ exclusive photoproduction
may proceed through open-charm exchange, namely $\Lambda_c\bar{D}^{(*)}$.
The authors point out that the thresholds for these intermediate states are very close
to the $J/\psi $ threshold and their exchange can contribute to the reaction.
They predict cusps in the total cross section at the $\Lambda_c\bar{D}$ and 
$\Lambda_c\bar{D}^{*}$ thresholds. 
If such a mechanism would dominate over the
gluon-exchange mechanism, it would obscure the relation between $J/\psi $
exclusive photoproduction and the gluonic properties of the proton
together with all the important physical implications discussed above.

\begin{figure}[b]
\includegraphics[width=0.45\textwidth]{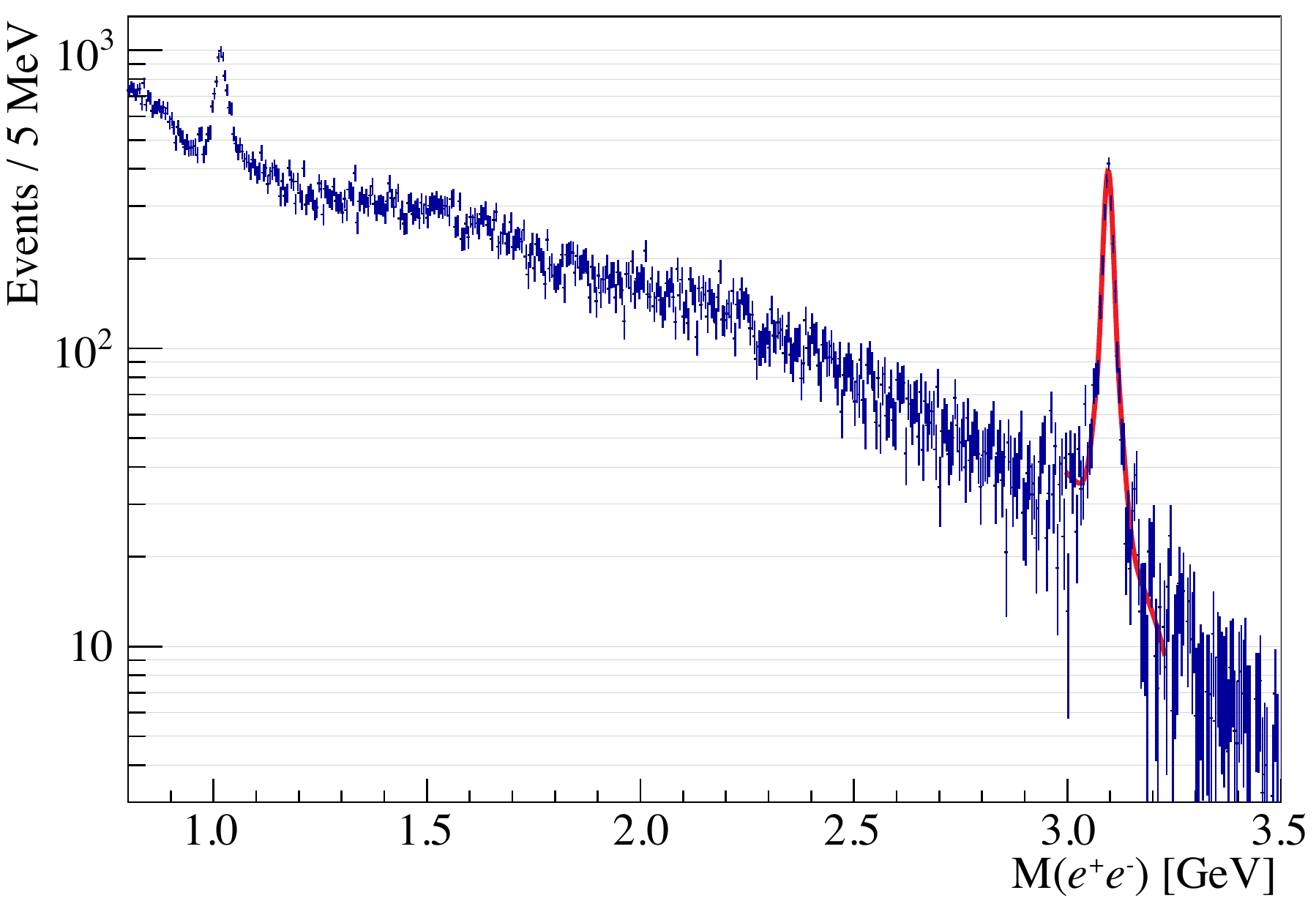}
  \caption{
The $e^+e^-$ invariant mass spectrum
for the GlueX Phase-I data set after applying the selections described in Section \ref{sec:analysis}.
The $J/\psi $ peak is fitted with a linear function and two Gaussians with common mean, which yields a total of $2270 \pm 58$ $J/\psi $'s.
} 
  \label{fig:minv_all}
\end{figure}

Furthermore, understanding the contribution of any processes besides
gluon exchange to $J/\psi$ photoproduction is crucial for the search
for the photoproduction of the LHCb $P_c^+$ pentaquark candidates~\cite{LHCb1,LHCb3}.  The
$P_c^+$ states can be produced in the $s$-channel of the $\gamma p \to J/\psi p$ 
reaction, and the strength of this resonant contribution can
be related to the branching fraction of $P_c^+\to J/\psi p$ under the
assumptions of VMD and a dominant non-resonant gluon exchange~\cite{Wang, Kubarovsky, Karliner, Blin}. If
there would be significant contributions from other processes such as the
open-charm exchange mentioned above, both of these assumptions break
down. Therefore, a better understanding of all the processes
that contribute to $J/\psi$ photoproduction is required before updated
searches for the $P_c^+$ can be performed.

In this work we report on the measurement of $J/\psi $ exclusive photoproduction,
\begin{equation}
\label{eq:jpsi_reaction}
\gamma p \to J/\psi p \to  e^+e^-p \;,
\end{equation}
based on the data collected by Phase-I of the GlueX experiment~\cite{GlueX_NIM} during the period $2016-2018$.
This data sample is more than four times larger than the one used in the first GlueX publication~\cite{prl_gluex}.
We present results for the total cross section
for photon beam energies from threshold, $E_{\gamma} =8.2$, up to $11.4$~GeV.
We also present the differential cross sections, $d\sigma/dt$,  
in three regions of photon beam energy over the full kinematic space in momentum transfer $t$, from 
$|t|_\mathrm{min}(E_\gamma)$ to $|t|_\mathrm{max}(E_\gamma)$,
thanks to the full acceptance of the GlueX detector for this reaction.
We identify the $J/\psi $ particle through its decay into an electron-positron pair. 
Due to the wide acceptance for the exclusive reaction $\gamma p \to e^+e^-p$, 
we observe events in a broad range of $e^+e^-$ invariant masses, including peaks corresponding 
to the $\phi$ and $J/\psi$ mesons
and the continuum between the two peaks that is 
dominated by the non-resonant Bethe-Heitler (BH) process 
(see Fig.~\ref{fig:minv_all}).
As an electromagnetic process that is calculable to a high accuracy, we will use the measurement of this BH process for the absolute normalization of the $J/\psi$ photoproduction cross sections.

\section{The GlueX detector}
\begin{figure}[b]
\includegraphics[width=0.4\textwidth]{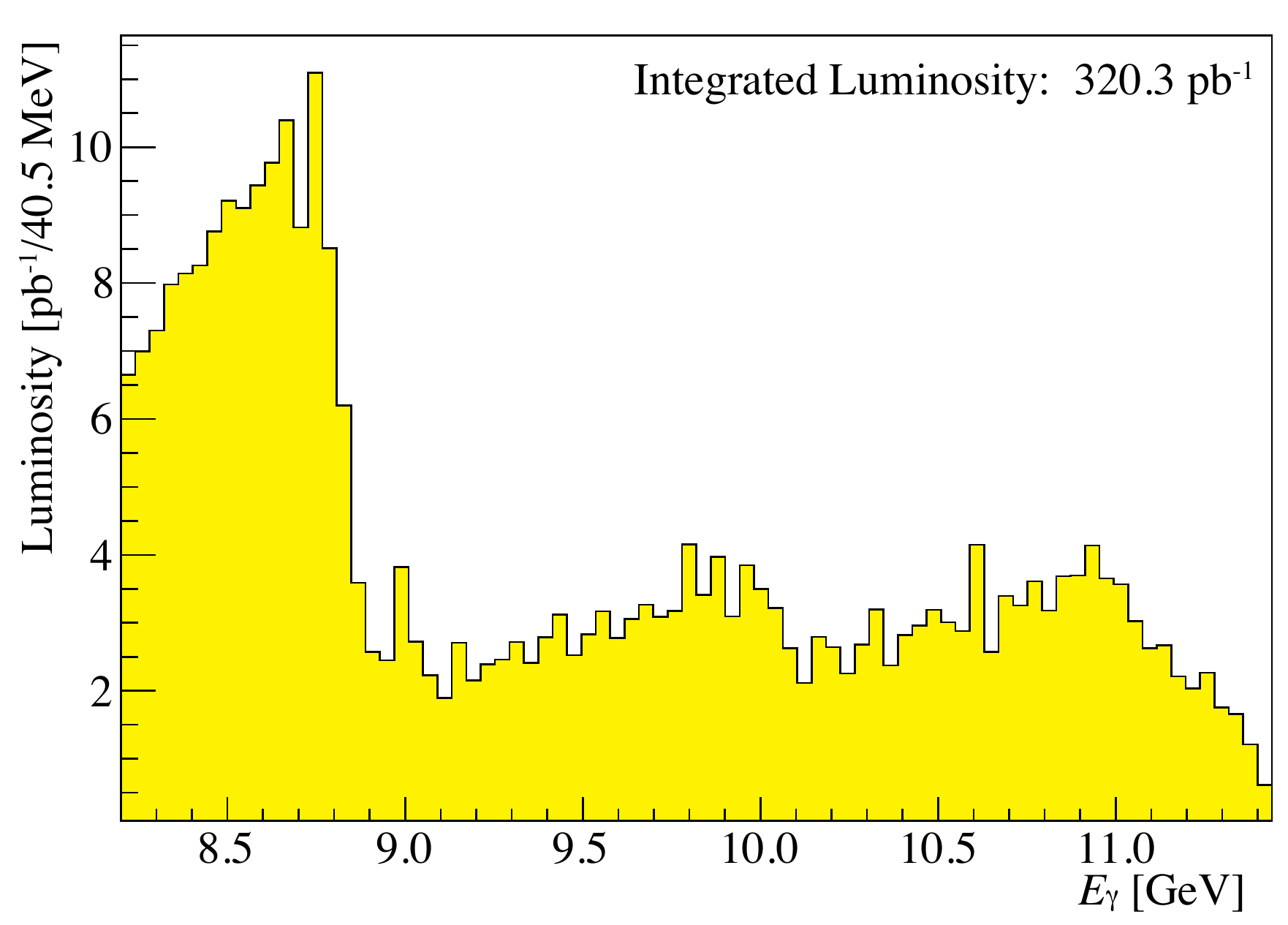}
  \caption{
The measured tagged photon spectrum for GlueX Phase-I in units of luminosity.
The non-statistical fluctuations are due to the segmentation of the tagger.
} 
  \label{fig:lumi}
\end{figure}

The experimental setup is described in detail in Ref.~\cite{GlueX_NIM}.
The GlueX experiment uses a tagged photon beam, produced on a diamond radiator 
from coherent Bremsstrahlung of the initial electron beam from the CEBAF accelerator.
The scattered electrons are deflected by a $9$~T$\cdot $m dipole magnet and 
detected in a tagging array which consists of scintillator 
paddles and fibers, 
that allows determination of the photon energy with $0.2\%$ resolution. 
The photons are collimated by a 5~mm diameter hole placed at $75$~m downstream of the radiator.
The flux of the photon beam is measured with a pair spectrometer (PS)~\cite{PS} 
downstream of the collimator,
which detects electron-positron pairs produced in a thin converter.
For most of Phase-I, the electron beam energy was $11.7$~GeV, 
corresponding to about $11.4$~GeV maximum tagged photon energy.
The coherent peak was kept in the region of $8.2-9.0$~GeV, which is just above
the $J/\psi $ threshold, see Fig.~\ref{fig:lumi}.
The produced photon beam is substantially linearly polarized in this peak region and the orientation
of the polarization was changed periodically, although the beam polarization was not used in this analysis. 
The bunches ($\approx 1$~ps long) in the electron and secondary photon beams 
are $4$~ns apart for almost all of the data.

The GlueX detector is built around a $2$~T solenoid, which is 4~m long 
and has an inner diameter of the bore of $1.85$~m.
A liquid Hydrogen target that is $30$~cm long, is placed inside the magnet.
It is surrounded by a Start Counter~\cite{ST}, a segmented scintillating detector
with a timing resolution of $250$ to $300$~ps, that helps us to choose the correct beam bunch.
The tracks of the final state charged particles are reconstructed using two drift chamber
systems. 
The Central Drift Chamber (CDC)~\cite{CDC} surrounds the target and consists
of 28 layers of straw tubes (about $3500$ in total) with axial and stereo orientations.
The low amount of material in the CDC allows tracking of the recoil protons down to momenta $p_p$
as low as $0.25$~GeV and identify them via the energy losses for $p_p<1$~GeV.
In the forward direction, but still inside the solenoid, 
the Forward Drift Chamber (FDC)~\cite{FDC} system is used to track charged particles.
It consists of 24 planes of drift chambers grouped in four packages
with both wire and cathode-strip (on both sides of the wire plane) readouts, 
in total more than $14,000$ channels.
Such geometry allows reconstruction of space points in each plane
and separation of trajectories in the case of high particle fluxes present
in the forward direction.

Electrons and positrons are identified by two electromagnetic calorimeters.
The Barrel Calorimeter (BCAL)~\cite{BCAL} is inside the magnet and surrounds the two
drift chamber systems.
It consists of lead layers and scintillating fibers, 
grouped in 192 azimuthal segments and four radial layers,
allowing reconstruction of the longitudinal and transverse shower development.
The Forward Calorimeter (FCAL) covers the downstream side of the acceptance
outside of the magnet at about $6$~m from the target and consists of 2800 lead-glass
blocks of ($4\times 4\times 45$)~cm$^3$.
A Time-of-Flight scintillator wall is placed just upstream of the FCAL.

The two calorimeters, BCAL and FCAL, are used to trigger the detector readout 
with a requirement of sufficient total energy deposition.
The trigger threshold is optimized for the collection of minimum ionizing events and is much lower than the sum of the energy
of the two leptons for the reactions discussed in this paper.
The intensity of the beam in the energy region above the $J/\psi $ threshold 
gradually increased from about $2\times 10^7$ photons/s in 2016
to about $10^8$ photons/s at the end of 2018,
resulting in a total integrated luminosity of $320$~pb$^{-1}$.

\section{Data Analysis}
\label{sec:analysis}
A key feature of our measurement is that the GlueX detector has essentially full acceptance 
for the $J/\psi$ photoproduction in Eq.~\ref{eq:jpsi_reaction}.
For photoproduction of light mesons, the acceptance of the recoil proton is limited at low momentum where the protons do not reach the drift chambers. 
However, due to the high mass of the $J/\psi $ meson, the recoil proton has a minimum
momentum of $0.6$~GeV and can be reliably detected.
Geometrically, the GlueX detector has full azimuthal acceptance 
and $1^\circ-120^\circ$ polar angle coverage, 
allowing detection of all the final state 
particles in the whole kinematic region of the reaction.
Thus, the total cross section of the exclusive reaction is measured directly,
without any assumptions about the final state particles or
extrapolations to kinematic regions outside of the acceptance.

The three final state particles are required to originate within the time of the same
beam bunch. 
The beam photons whose time (as determined by the tagger) coincide with this bunch
are called in-time photons 
and they qualify as candidates associated with this event. 
The other, out-of-time, tagged photons are used to estimate the fraction of events that are ``accidentally''
associated with an in-time photon that did not produce the reconstructed final state particles.
Unless otherwise noted, all the distributions shown in this paper have 
the corresponding accidental background contributions subtracted. 

The exclusivity of the measurement, 
together with the precise knowledge of the beam energy and its direction, 
allows performing of a kinematic fit.
The fit requires four-momentum conservation and a common vertex 
of the final state particles.
A very loose selection criterion is applied to the $\chi^2$ value of the fit.
The momentum of the recoil proton, $p_p$ is relatively well measured, 
as the protons are produced at moderate polar angles ($\theta\approx10-30^\circ$) with $p_p \approx 1$~GeV.
This is not the case for the lepton pair, where
one of the leptons is predominantly produced with a high momentum at a small polar angle,
i.e. in a region with 
a poor momentum resolution of the solenoidal spectrometer.
The kinematic fit to the full reaction is therefore constrained mainly by the direction and magnitude
of the proton momentum and the direction of the lepton momenta, which are measured 
more precisely than the magnitudes of the lepton momenta.
After applying the kinematic fit, the 
 $J/\psi $ mass resolution improves significantly
to about $13$~MeV
(see Fig.~\ref{fig:minv_all}).

Monte Carlo simulations for both $J/\psi $ and BH processes have been performed.
To calculate the absolute BH cross section, we have used a generator~\cite{Rafo} based on
analytic calculations of the BH cross sections~\cite{Berger}. 
For the proton form factors that enter in the calculations,
we use the low-$Q^2$ parametrization of Ref.~\cite{protonFF}.
We note that if the dipole form factors are used instead, the BH cross section differs by less than 1\% 
within the kinematic region used for normalization.
The $J/\psi $ events were generated using a $t$-dependence 
and an energy dependence of the cross section 
obtained from smooth fits to our measurements. 
For the $J/\psi $ decay, photon-to-$J/\psi $ spin projection conservation 
in the Gottfried-Jackson frame is assumed.
This corresponds to a $1+\cos^2 \theta_\mathrm{GJ}$ angular distribution of the 
decay particles, where $\theta_\mathrm{GJ}$ is the lepton polar angle
in the Gottfried-Jackson frame.

To simulate the detector response we have used the GEANT4 package~\cite{GEANT4}.
In addition, to the generated events, we include  
accidental tagger signals and detector noise hits
extracted from data collected with an asynchronous trigger.
These simulations are used to calculate the reconstruction efficiencies 
for the two processes, $\varepsilon _\mathrm{BH}$ and  $\varepsilon _{J/\psi }$.
The BH simulations are also used to integrate the absolute
cross sections in the kinematic regions used for normalization.

\label{sec:BH}
We use the BH process 
in the $e^+e^-$ invariant mass region of $1.2<M(e^+e^-)<2.5$~GeV for
the absolute normalization of the $J/\psi $ total cross section,
thus eliminating uncertainties 
from sources like luminosity and reconstruction efficiencies that are common 
for both processes.
The main challenge in extracting the BH yields is to separate the pure $e^+e^-$ 
continuum from the background of $\pi^+\pi^-$ production that is more than three orders of magnitude
more abundant. We suppress the pions primarily using the energy deposition $E$
in the calorimeters and requiring both lepton candidates
to have $p/E$ consistent with unity, where $p$ is the momentum determined from the kinematic fit.
In addition, we use the inner layer of the BCAL as a pre-shower detector and require
the energy deposition there to be $E_\mathrm{pre}\,\sin\theta > 30$~MeV,
where $\sin\theta $ corrects for the path length
in the pre-shower layer.
The pion background is further reduced by selecting the kinematic region 
with particle momenta $p>0.4$~GeV, to remove pions coming from target excitations.
In addition, for the BH measurements only, we select $|t|<0.6$~GeV$^2$ as
the BH cross section is dominated by the pion background above this $t$-value, due to the very sharp $t$-dependence of the BH process.
After applying all of the selection criteria above, the remaining background 
is of approximately the same magnitude as the signal.
The final BH yields are extracted by subtracting this pion background using the procedure described below.

\begin{figure}
    \includegraphics[width=0.35\textwidth]{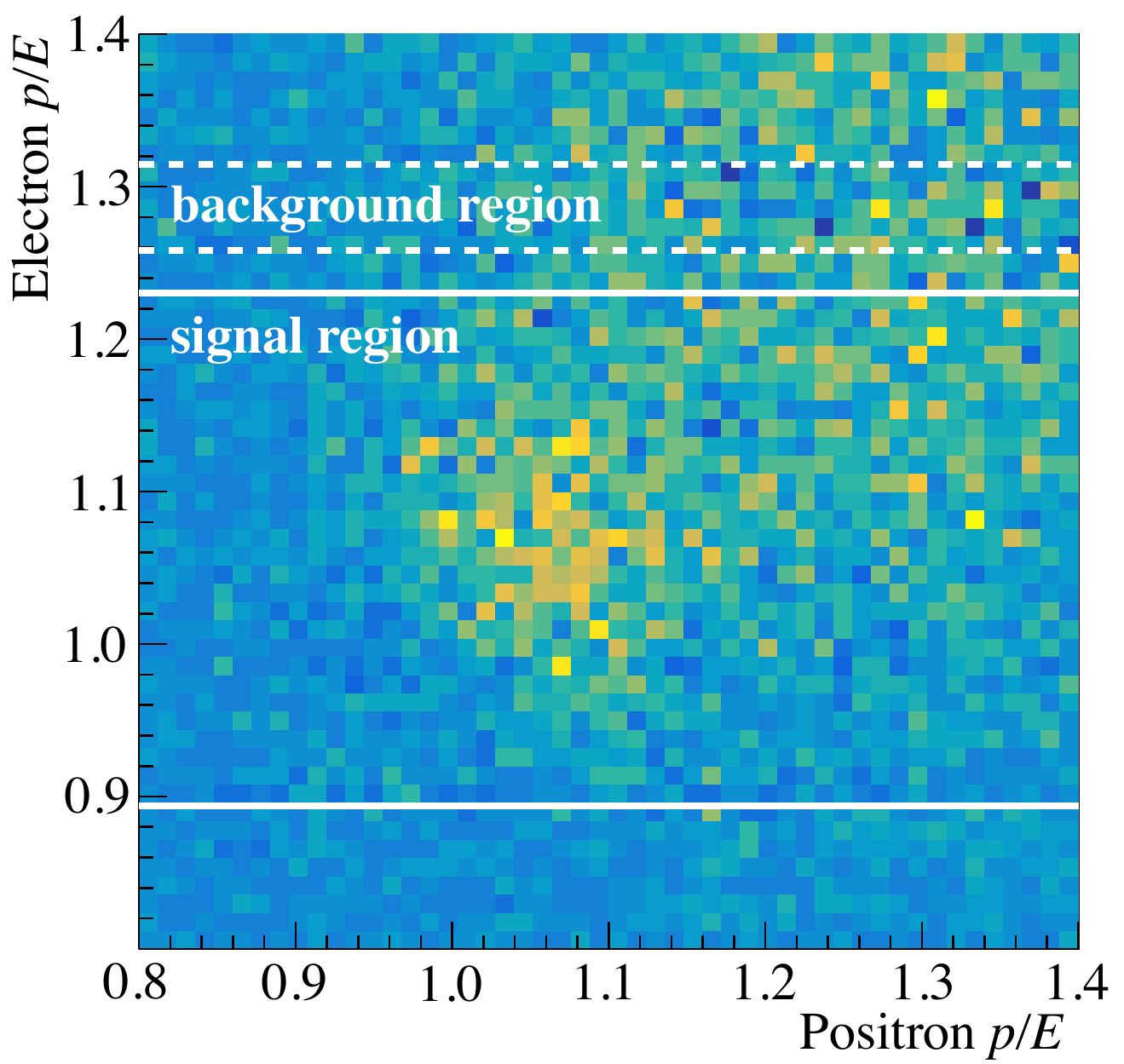}
  \caption{
Electron vs. positron $p/E$ distribution in the BH invariant mass region of $1.2-2.5$~GeV.
The white horizontal lines indicate the background and signal regions used when projecting onto the positron axis.
See text for explanations.
}
    \label{fig:epem2d_bh}
\end{figure}
\begin{figure}
    \includegraphics[width=0.5\textwidth]{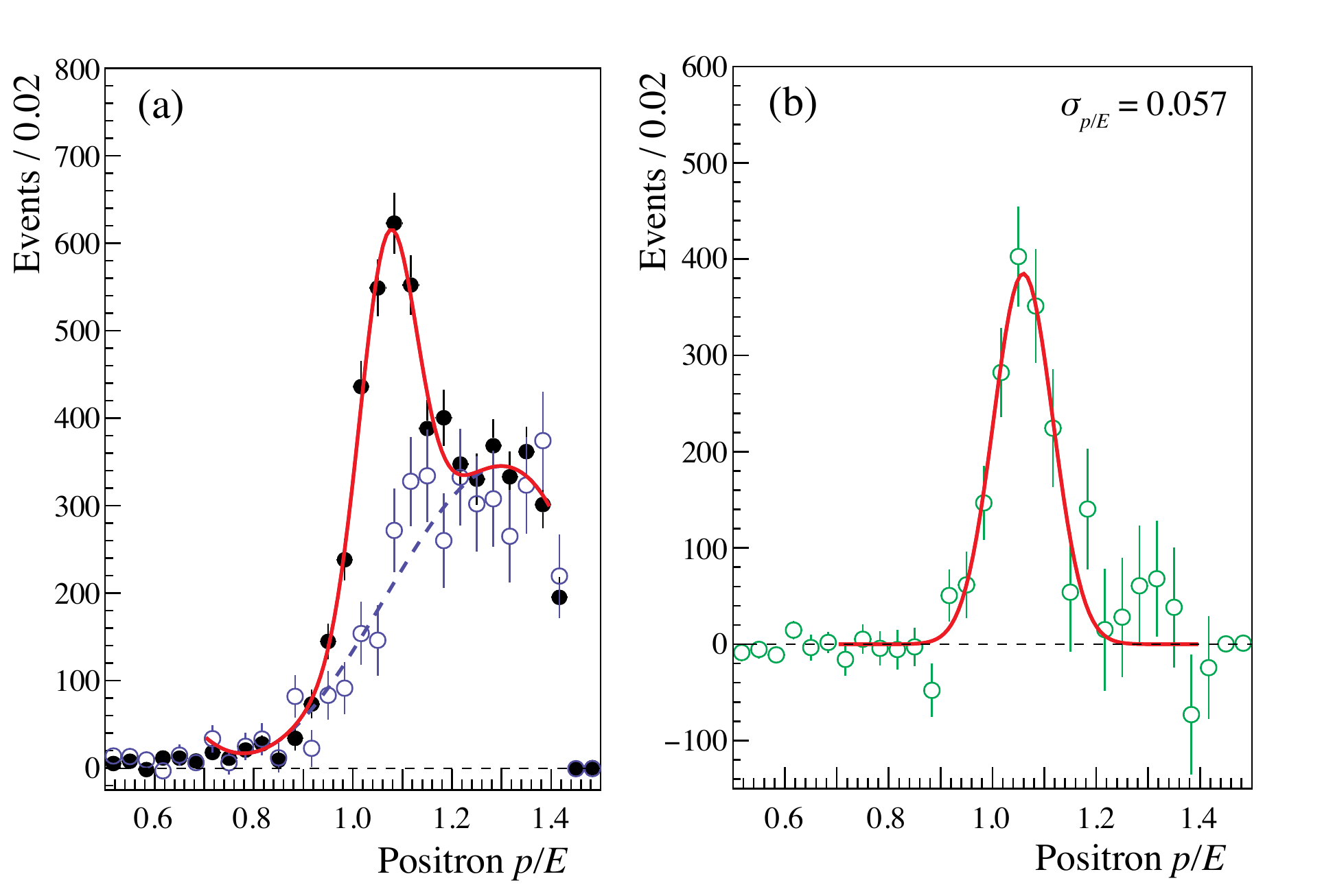}
  \caption{
(a) The $p/E$ distributions in the signal (solid black) and background (open blue) region
 for the BH
invariant mass region of $1.2-2.5$~GeV. 
(b) The difference between the signal  
and background distributions from (a).
See text for explanation of the fits.
}
    \label{fig:BFCALres_bh}
\end{figure}

We extract the yields of the leptons detected in the BCAL and FCAL separately, 
since the calorimeters have different resolutions.
We perform this procedure in bins of the beam energy or other 
kinematic variables.
For illustration only, in Figs.~\ref{fig:epem2d_bh} and \ref{fig:BFCALres_bh} 
we demonstrate this procedure over one energy bin ($8.92< E_\gamma < 9.10$~GeV) 
including leptons detected in both calorimeters. 
We consider the two-dimensional $p/E$ distribution of electron vs. positron candidates, 
and define a one-dimensional $\pm 3\sigma_{p/E} $ signal region around the $p/E$ peak of one of the leptons.
The projection of this region onto the $p/E$ axis of the second lepton is shown in Fig.~\ref{fig:BFCALres_bh}(a) (full black points).
The shape of the pion background is estimated using events outside of the $p/E$ peak 
of the first lepton (the background region indicated in Fig.~\ref{fig:epem2d_bh}), 
to which we fit a polynomial function of third order. 
The events in the signal region (full black points in Fig.~\ref{fig:BFCALres_bh}(a)) 
are then fit with a sum of a Gaussian and this polynomial, where the latter is
multiplied by a free normalization parameter, $B_\mathrm{norm}$.
The background distribution scaled by $B_\mathrm{norm}$ is shown by the open blue
points in Fig.~\ref{fig:BFCALres_bh}(a).
The lepton yields are extracted by fitting the difference 
of the distribution in the signal region (full black points)
and the scaled background distribution (open blue points) with a Gaussian,
shown in Fig.~\ref{fig:BFCALres_bh}(b).
We perform this procedure for both positrons and electrons.
For each species, the yields are extracted separately for the cases 
where the selected lepton is detected by the BCAL or the FCAL 
(regardless of where the other lepton is detected). 
We then average the summed yields for electrons and positrons to estimate the BH yields.
To estimate the systematic uncertainty of this procedure at each data point, 
two variations of the method are tested.
They differ by fixing the width of the $p/E$ peak to the simulations 
(default for the central value)
or leaving it as a free parameter. 
We also vary the method of integrating the signal, either by summing the histogram
values in Fig.~\ref{fig:BFCALres_bh}(b) (default) or integrating the fitted function. 
The results of these variations
are discussed in Sec.~\ref{sec:ptp_syst}.

As a check of the validity of our reconstruction procedure, 
we extract the BH cross section from our data 
and compare it to the expectations from the  
absolute calculations described previously.
The fitting procedure described above is applied in bins
of various kinematic quantities, e.g. $E_\gamma $, and 
we extract the cross section as: 
\begin{eqnarray}
\begin{array}{l}
\sigma_\mathrm{BH}^\mathrm{data}(E_\gamma ) = \frac{N_\mathrm{BH}(E_\gamma )}{L(E_\gamma )\varepsilon _\mathrm{BH}(E_\gamma )},
\label{eq:bhxsec_data}
\end{array}
\end{eqnarray}
where $N_\mathrm{BH}$ is the measured BH yield in a specific photon-energy bin, 
$\varepsilon_\mathrm{BH}$ is the corresponding reconstruction efficiency determined
from MC simulations, and $L$ is the measured luminosity.  
We note that the photon beam luminosity is used just for this study as a cross check, 
but not for the final $J/\psi$ cross sections 
that are determined relative to these BH cross sections.

\begin{figure}[h]
    \includegraphics[width=0.45\textwidth]{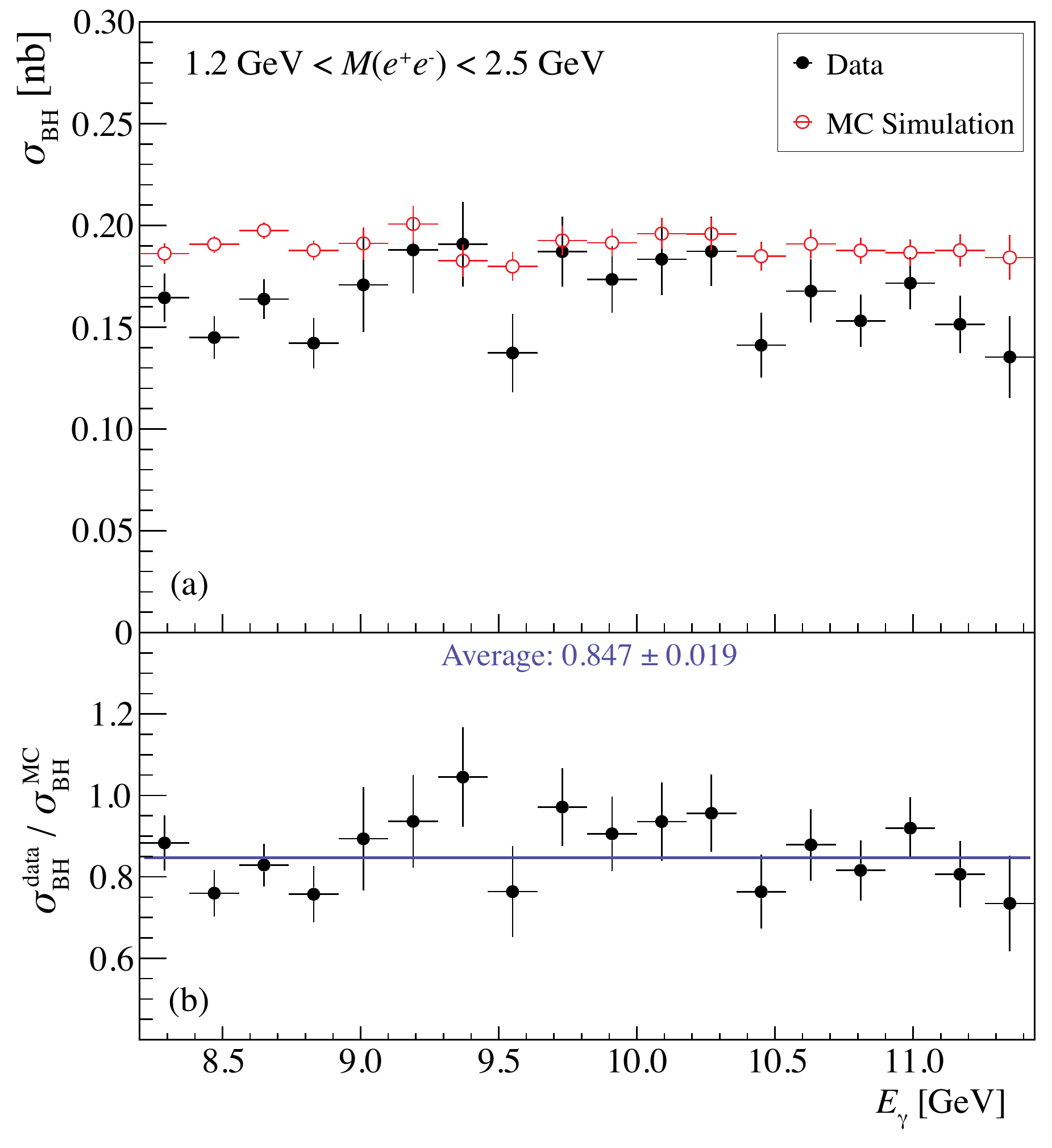}

\caption{BH cross section vs. beam photon energy for $1.2 < M(e^+e^-)<2.5$~GeV. (a) BH cross section obtained from data and MC simulation. (b) Ratio of data and MC cross sections from (a) fitted with a constant.}
\label{fig:BH_data_MC}
\end{figure}

\begin{figure}[h]
    \includegraphics[width=0.45\textwidth]{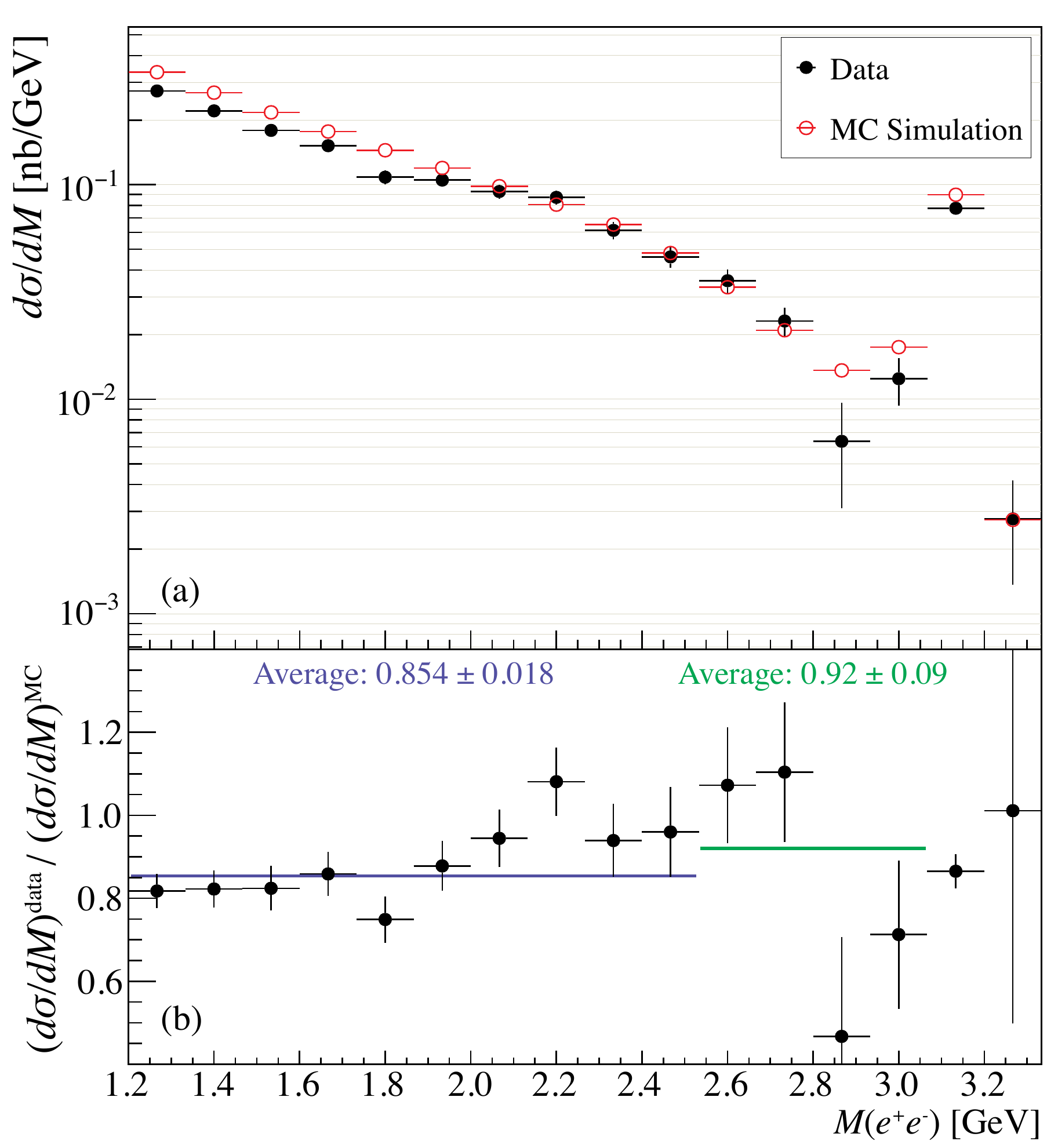}
  
\caption{The sum of BH and $J/\psi$ cross sections as function of $e^+e^-$ invariant mass. 
(a) Cross section obtained from data and MC simulation. 
(b) Ratio of data and MC cross sections from (a)
fitted with constants in two regions:
the region used for normalization of the $J/\psi$ cross section, and
 the vicinity of the $J/\psi $ peak.}
\label{fig:BH_Minv_data_MC}
\end{figure}

The BH cross sections as function of the beam energy, extracted from
Eq.~(\ref{eq:bhxsec_data}), 
are compared with the MC calculations in Fig.~\ref{fig:BH_data_MC}.
The data/MC ratio of the cross sections 
(Fig.~\ref{fig:BH_data_MC}(b)) is consistent with a constant 
and differs from unity by about $15\%$.
Since this ratio is approximately constant over the kinematic region under consideration, we take its difference from unity as an estimation of the overall systematic uncertainty in the normalization of our cross sections.
Similar ratios as a function of other kinematic variables,  including proton momentum and polar angle,
have been studied.  
Although the BH cross section varies by up to two orders of magnitude across these variables, 
the data and MC results remain consistent.
In Fig.~\ref{fig:BH_Minv_data_MC} we show one such comparison as a function of the invariant mass, $M(e^+e^-)$, which illustrates
how well the BH simulations describe the data from the region used for normalization ($1.2-2.5$~GeV) to the $J/\psi $ peak.
We see a slight increase in the data/MC ratio 
in the region close to the $J/\psi$ peak, 
which, however, is not statistically significant and 
is within the $15\%$ uncertainty estimated above.

\label{sec:JP}
To measure the $J/\psi$ yields, we apply the same event selections as for the BH process 
described above, except that we do not constrain the $|t|$ range.  
We select lepton candidates using $\pm 3\sigma $ $p/E$ selections, 
however, in contrast to the BH continuum,
no additional $p/E$ fitting procedure is needed to separate the pion background.
Instead, we separate $J/\psi$ candidates from the background by fitting the narrow $J/\psi$ peak
in the $M(e^+e^-)$ distributions.
We fit the mass distributions in 18 bins of the beam energy with a Gaussian 
for the $J/\psi $ peak plus a linear background.
Because of the fine binning and the resulting small sample size in each bin,
we employ the binned maximum-likelihood
method, where Poisson errors are assumed in each invariant-mass bin, 
using the {\it RooFit} package~\cite{RooFit}. 
Our studies show that the background due to accidental beam photon combinations 
in this mass region is small (about $5\%$) and of similar shape to the other smooth 
backgrounds, so in this case we do not explicitly subtract these accidental combinations.
We perform fits, where we leave the Gaussian width of the $J/\psi $ peak as a free parameter 
and where we fix it to the expectation from MC simulation.
The fitted widths of the $J/\psi $ peaks match well the expectations from simulation. 
We hence fix the widths to obtain our nominal results and use the results with free widths 
to estimate the systematic uncertainty in our knowledge of the peak shape.
To study the systematic uncertainty of the lepton identification we also vary the $p/E$ selections, 
and include these variations as described below.

\section{Total cross section}
\begin{figure}[]
\includegraphics[width=0.45\textwidth]{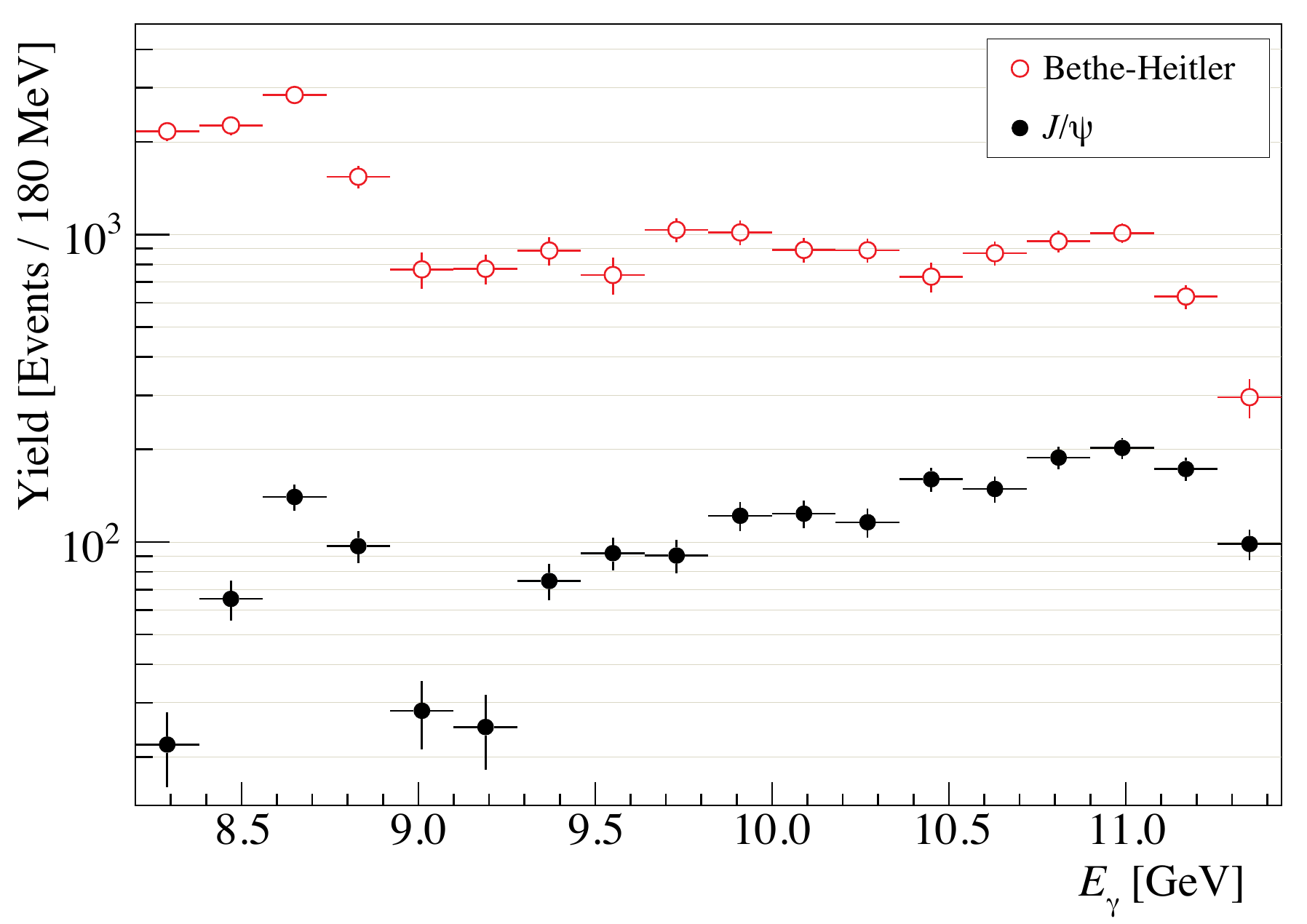}
  \caption{
Comparison of BH and $J/\psi $ yields as function of beam energy.
} 
  \label{fig:yield_BH_JP}
\end{figure}
\begin{figure*}
\includegraphics[width=\textwidth]{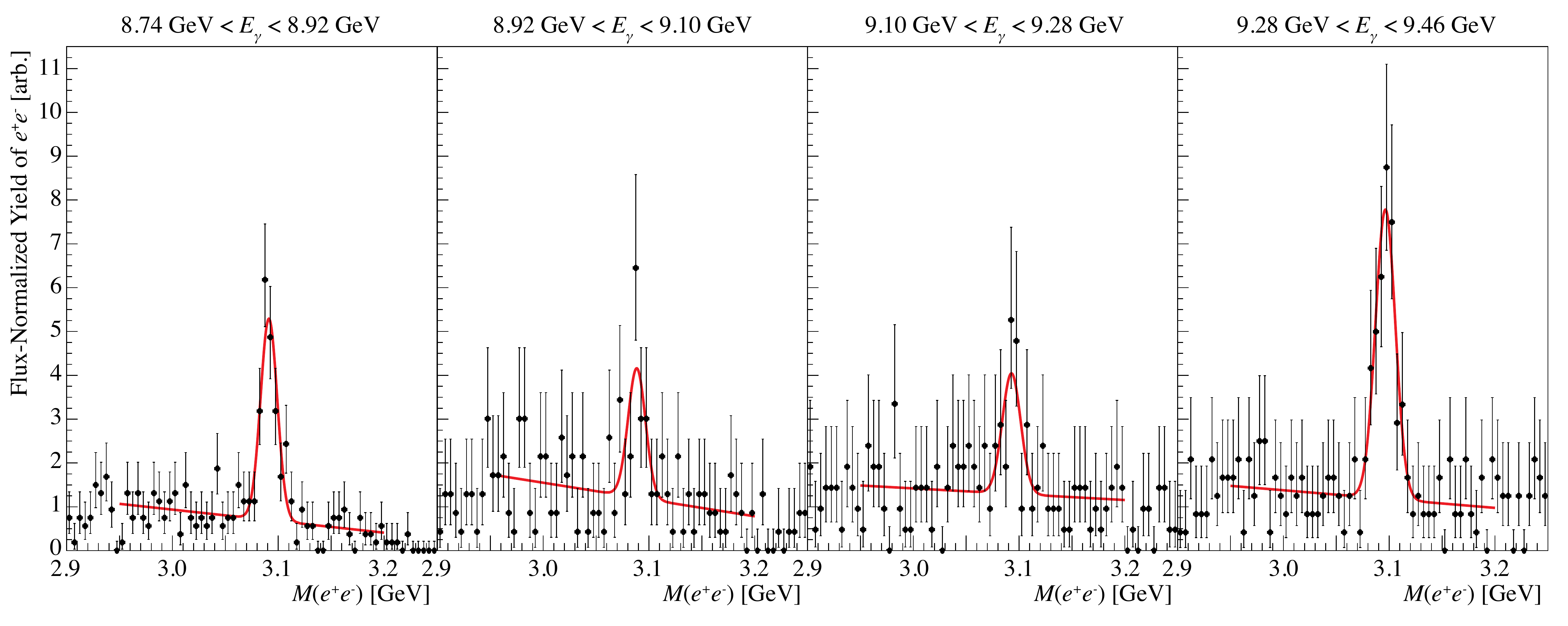}
\caption{
Distribution of $M(e^+e^-)$ in bins of beam energy $E_\gamma$ with fits to the $J/\psi$ peak overlaid.
The $y$ axis of each histogram is scaled by the flux integrated over the corresponding $E_\gamma$ bin.
}\label{fig:JP_fits}
\end{figure*}

The extracted $J/\psi $ and BH yields as a function of beam energy
are compared in Fig.~\ref{fig:yield_BH_JP}.
While the BH yields follow the beam intensity spectrum,
the $J/\psi $ yields exhibit an indication of a dip in the $9.1$~GeV region
which will be discussed below.
For illustration,
individual $J/\psi $ mass fits for four energy bins around $9.1$~GeV  
are shown in Fig.~\ref{fig:JP_fits}. 
The beam photon flux varies strongly in this region, so to
correct for this effect we scale the yield by the flux for the
corresponding energy bin.

\label{sec:jp_fits}

We calculate the total cross section as a function of beam energy using the following formula:
\begin{eqnarray}
\begin{array}{l}
\sigma(E_\gamma) = \frac{N_{J/\psi }(E_\gamma)}{N_\mathrm{BH}(E_\gamma)}~\frac{\sigma_\mathrm{BH}(E_\gamma)}{\mathrm{BR}_{J/\psi}}~ \frac{\varepsilon_\mathrm{BH}(E_\gamma)}{\varepsilon_{J/\psi }(E_\gamma)}.
\label{eq:xsec}
\end{array}
\end{eqnarray}
Here $N_{J/\psi}$ and $N_\mathrm{BH}$ are the corresponding yields, 
$\sigma_\mathrm{BH} $ is the calculated BH cross section 
integrated over the region used for normalization,
$\mathrm{BR}_{J/\psi}$ is the $J/\psi \rightarrow e^+e^-$ 
branching ratio of $5.97\%$~\cite{pdg}, and
$\varepsilon_{J/\psi }$ and $\varepsilon_\mathrm{BH}$ are the MC-determined efficiencies.  
Note that only the relative efficiency between the two processes
enters in the above equation.

\begin{figure}[]
\includegraphics[width=0.45\textwidth]{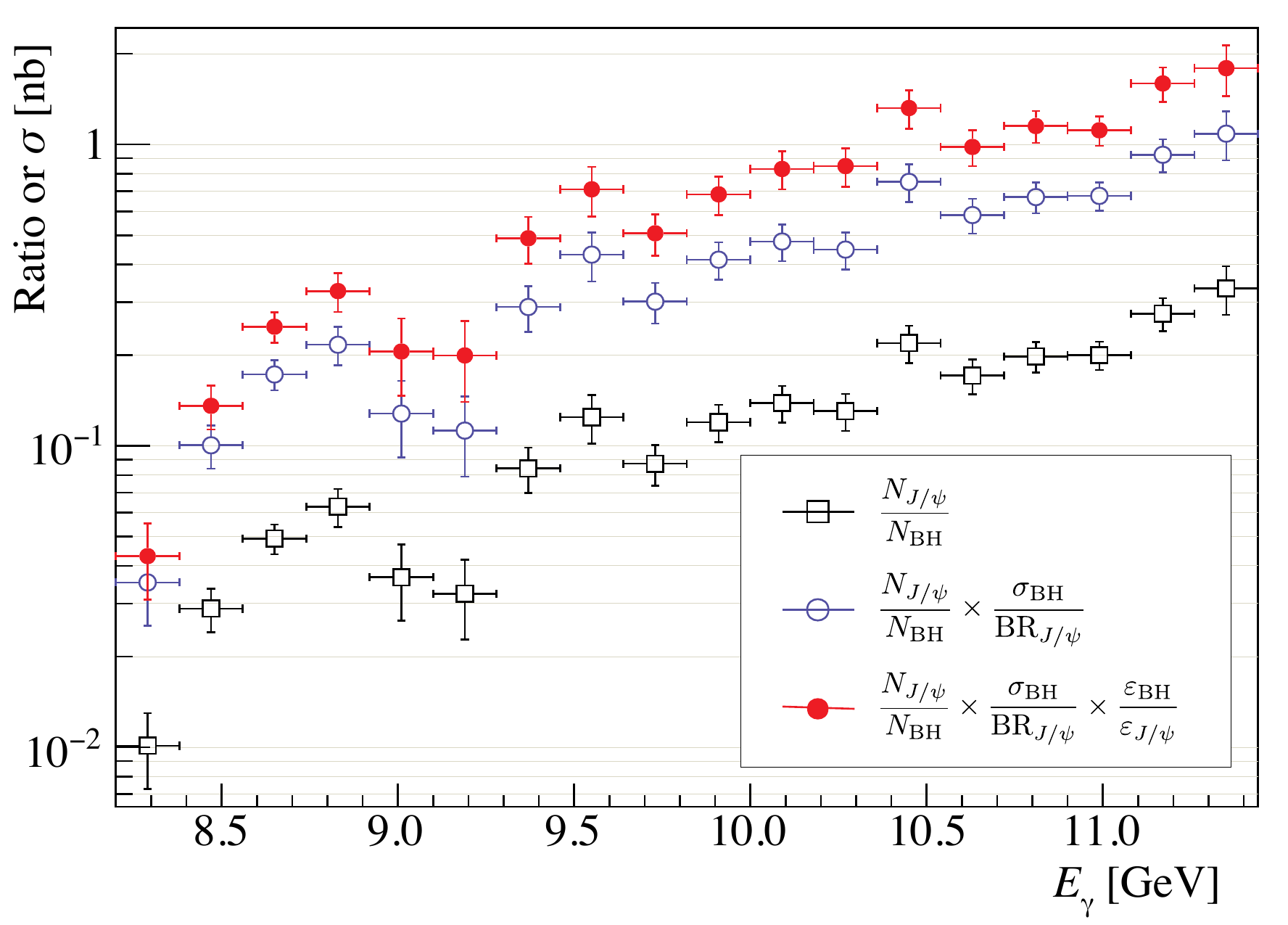}
  \caption{
Intermediate results of the $J/\psi $ cross-section calculation:
the $J/\psi $ to BH yield ratio (black open squares); 
the yield ratio multiplied by the BH cross section over $BR(J/\psi \to e^+e^-)$ (blue open points);
 and this result further
corrected by the BH-to-$J/\psi $ efficiency ratio (solid red points) corresponding to the final cross section (See Eq.(\ref{eq:xsec})).
The nb units are valid for the final result only (solid red points).
Only the statistical errors are shown.
} 
  \label{fig:xsec_evol}
\end{figure}

The calculations in Eq.(\ref{eq:xsec})
are shown in several steps in Fig.~\ref{fig:xsec_evol} to demonstrate that 
the possible dip structure at $E_\gamma \approx 9.1$~GeV 
arises from the yield ratio and not from the subsequent corrections.
We note that the position of the dip coincides with a drop 
in the photon-beam intensity just above the coherent peak, as seen in Fig.~\ref{fig:lumi}, however we perfomed studies showing that this is coincidental.
In particular, as seen in  Fig.~\ref{fig:yield_BH_JP}, there is no dip in the BH yields in this region.
Since the reconstruction of the $e^+e^-p$ final state is strongly determined by the reconstruction of the recoil proton, 
we have also searched for a similar deviation in $p\bar{p}$ photoproduction, $\gamma p \to (p\bar{p})p$,
where we require the $p\bar{p}$ invariant mass to be in the $J/\psi $ mass region
$3.05<M(p\bar{p})<3.15$~GeV.
With this selection, the recoil protons in this reaction are kinematically close to
those in the $\gamma p \to J/\psi p$ reaction.
We find that the flux normalized yields for the $p\bar{p}$ reaction as a function of photon energy are smooth 
in the region of the $J/\psi $ dip.

\begin{figure}[]
\includegraphics[width=0.45\textwidth]{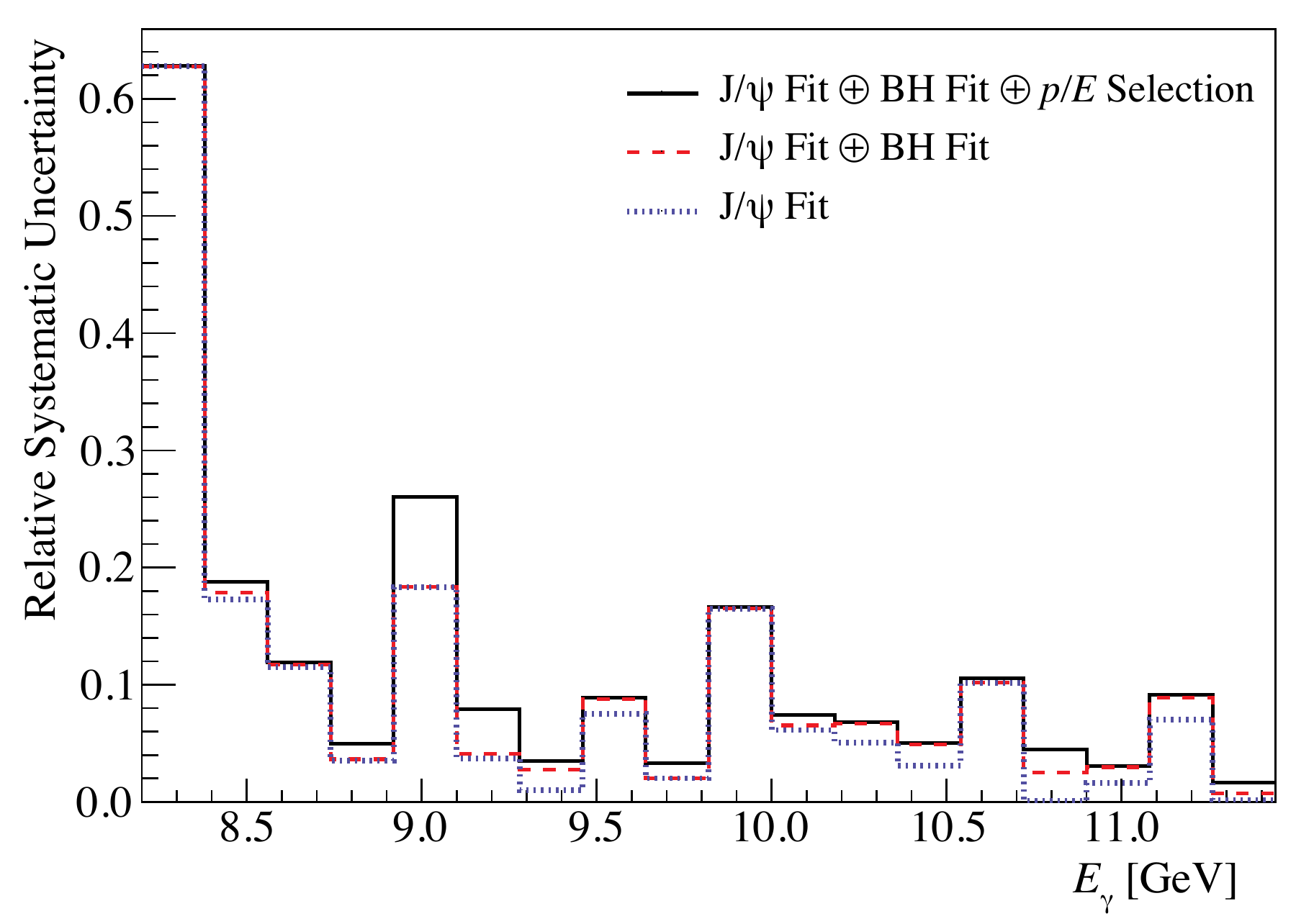}
  \caption{
Contributions of the different sources 
to the systematic uncertainties of the individual energy bins, successively added in quadrature.
} 
  \label{fig:Syst_ptp}
\end{figure}

\label{sec:ptp_syst}
The systematic uncertainties on the individual cross section points are taken from three sources as previously described.  
The systematic uncertainty in the BH yield extraction is determined by the maximum 
deviation in the two fitting variations from the nominal value, as discussed above. 
The systematic uncertainty in the $J/\psi$ yield extraction is determined by taking the difference
in the cross section values between the fits with fixed and free Gaussian widths.  
Additionally, we study the deviation in the cross section when widening
the selected $p/E$ region around the peak to $\pm 4\sigma$.
To estimate this uncertainty,
we use the photon flux instead of the BH cross section to calculate the cross section.
This change in the normalization is required due to the difficulty 
of measuring the BH cross section with this looser $E/p$ requirement.
The uncertainties from each of the three contributions are added in quadrature to get the total systematic uncertainties.  These values are illustrated in Fig.~\ref{fig:Syst_ptp}.

As mentioned above,
we assume in the MC simulation a certain angular distribution of the $J/\psi $ decay products,
namely $1+\cos^2 \theta_\mathrm{GJ}$, where $\theta_\mathrm{GJ}$ is the lepton polar angle
in the Gottfried-Jackson frame, which corresponds to photon-to-$J/\psi $
conservation of the spin projection in this frame.
To estimate the systematic error related to this assumption, we compare 
the efficiency from this model to the extreme case when assuming uniform 
distribution.
The variations of the efficiency as a function of energy do not exceed $5\%$.
We also perform a fit to the measured $\theta_\mathrm{GJ}$ distribution and find the results to be
consistent with the assumption of spin projection conservation,
which reduces the above upper limit on this uncertainty to a $<2\%$ level.

\begin{figure}[h]
\includegraphics[width=0.45\textwidth]{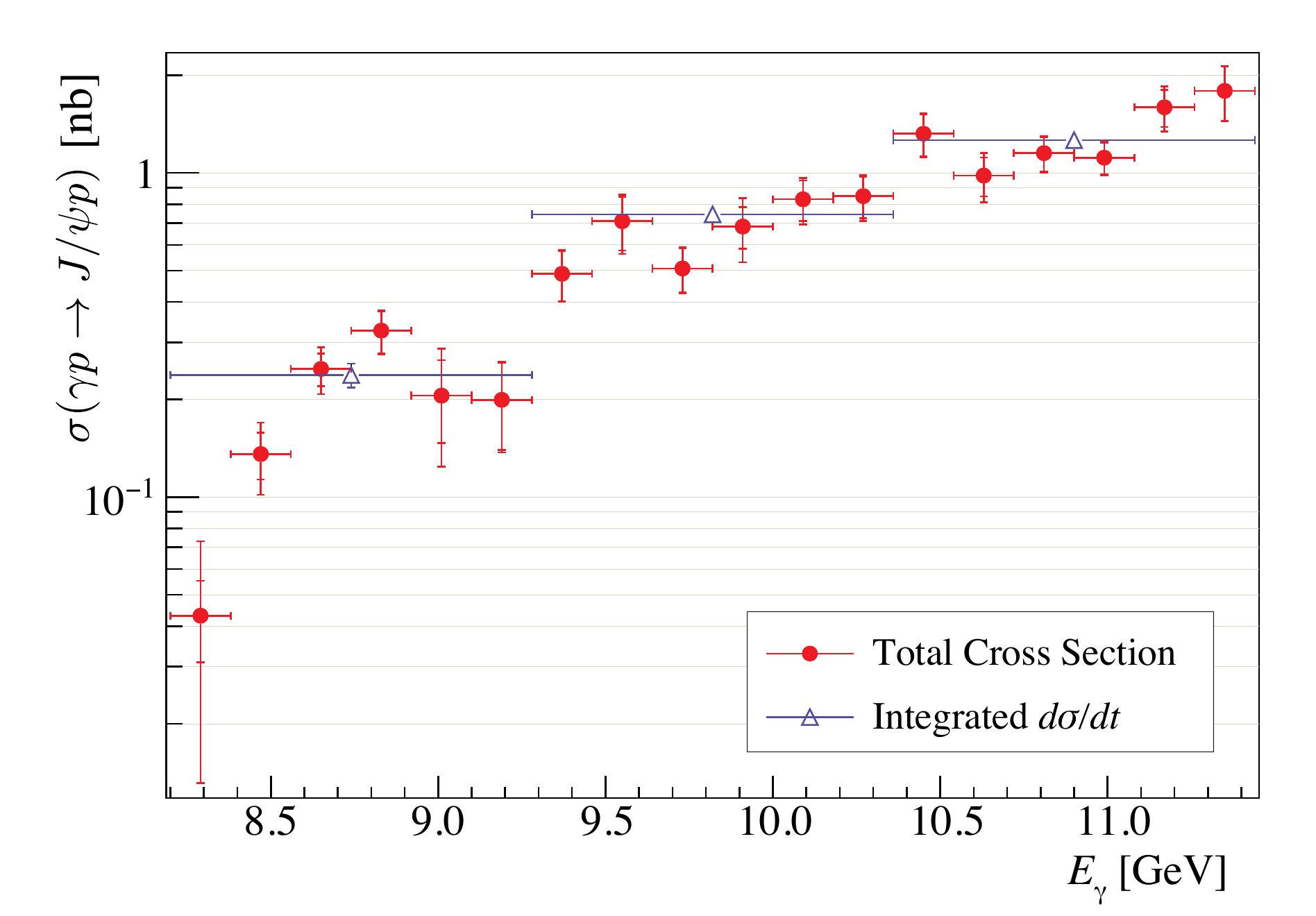}
  \caption{
The filled red points show 
the measured total cross sections obtained from Eq.(\ref{eq:xsec})
in fine photon energy bins. 
The inner bars represent the statistical errors and outer bars 
are the total errors, with the statistical and systematic errors added in quadrature. 
The open blue triangles represent
the total cross sections calculated by integrating the functions
fitted to the measured differential cross sections for the three beam energy regions, 
with only the statistical uncertainties shown.
}
  \label{fig:sigma_dsdt_comp}
\end{figure}

The measured total cross section is plotted in Fig.~\ref{fig:sigma_dsdt_comp}, 
 with the statistical and total uncertainties shown separately.
With the exception of the first point, the statistical errors dominate.
The numerical results for the total cross section, along with their statistical and systematic errors,
are given in Table~\ref{tab:sigma} of the Appendix, Sec.~\ref{sec:num_res}.

The summary of the sources and magnitudes of the overall normalization uncertainties is given in Table \ref{tab:total_syst}.
The main source of this uncertainty was discussed in Section~\ref{sec:BH},
where we studied the BH data/MC ratio as a function of the beam energy and invariant mass.
We use the $(15.3\pm1.9)\%$ average difference between data and MC, 
that is consistent with a constant function of energy (see Fig.\ref{fig:BH_data_MC}),
 as a measure of the systematic uncertainty in the overall scale.  
The effect of the radiative corrections to the cross section was studied
in the previous publication~\cite{prl_gluex} based on Ref.~\cite{Vanderhaeghen}.
The possible contribution of the Time-like Compton Scattering (TCS)
to the $e^+e^-$ continuum was estimated in ~\cite{prl_gluex} using a generator~\cite{Marie_note}
based on the calculations in Ref.\cite{Marie_journal}.
To estimate the effect of a possible contribution of $\rho^\prime (1600)$ 
to the $M(e^+e^-)$ region used for normalization, we fit the data/MC ratio vs. invariant mass in Fig.~\ref{fig:BH_Minv_data_MC}
with constants in two regions, the standard one $1.2-2.5$~GeV and the one over 
the $\rho '(1600)$ resonance region $1.46-1.86$~GeV.  The results are $0.854\pm0.018$ and $0.813\pm0.031$,
respectively. These results are consistent within the $3.6\%$ combined error, which we conservatively take as a measure of this systematic uncertainty.

\begin{table}[]
\caption{
Contributions to the overall normalization uncertainty and their sum in quadrature.
\label{tab:total_syst}
}\begin{ruledtabular}
\begin{tabular}{lcc}
\textrm{Source}&
\textrm{Uncertainty}\\
\colrule
BH data-to-MC ratio vs. $E_\gamma$ & $15.3\%$ \\
Radiative corrections  & $8.3\%$ \\
TCS contribution to BH& $8\%$ \\
$\rho^\prime$ contribution to BH  & $3.6\%$ \\
\colrule
Total &   $19.5\%$ \\
\end{tabular}
\end{ruledtabular}

\end{table}
\section{Differential cross sections}

\label{sec:dsdt_proc}
\begin{figure}[h]
\includegraphics[width=0.45\textwidth]{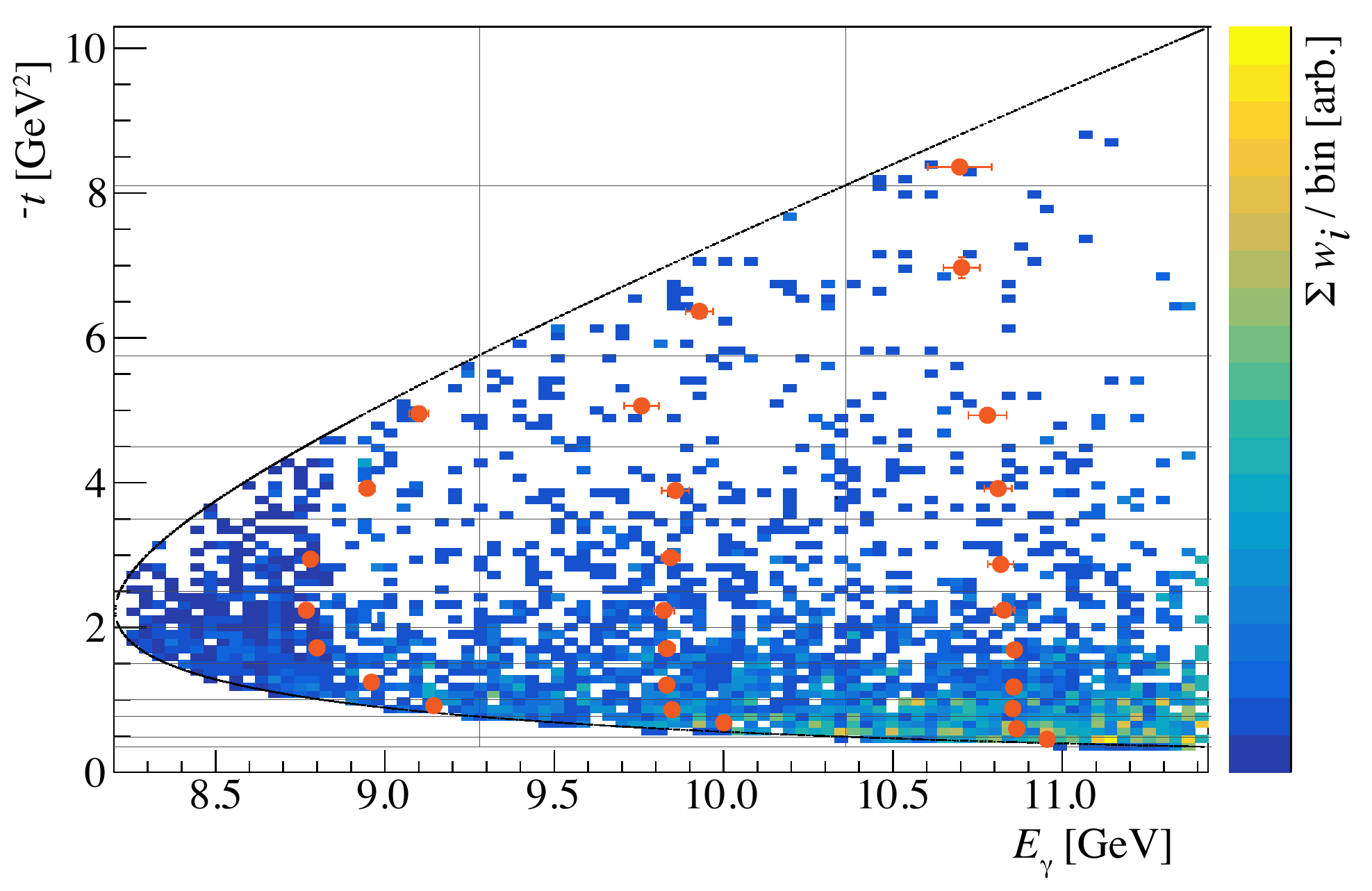}
  \caption{
The distribution of the flux-weighted data in the $E_\gamma-t$ plane and
the mean values of the reported cross sections (solid dots)
within the corresponding bins.
A mass selection of $3.05<M(e^{+}e^{-})<3.15$~GeV is used for the events in this plot. 
}
  \label{fig:dsdt_binning_data}
\end{figure}

We present measurements of the differential cross sections, $d\sigma/dt(E_\gamma,t)$,
over the entire near-threshold kinematic region. 
The two-dimensional bins in the ($E_\gamma,t$) plane for which we report the cross section values
are shown in Fig.~\ref{fig:dsdt_binning_data}.
We subdivide the data into three equidistant energy ranges, while the $t$-bins
match the crossing of these ranges with the $|t|_\mathrm{min}(E_\gamma)$ and $|t|_\mathrm{max}(E_\gamma)$
kinematic limits.
Such a choice allows sufficient sample size in each bin. 
Because the variation of the beam-photon flux across each energy bin is rather large,
we weight each event by the measured luminosity $L(E_\gamma)$ in steps of $45$~MeV bins,
i.e. the  weight for $E_\gamma $ bin $i$ is:
\begin{eqnarray}
\begin{array}{l}
{\mathrm{weight}}_i = \frac{1}{L(E_{\gamma i})[{\mathrm{nb}}^{-1}]/0.045{\mathrm{GeV}}}.
\label{eq:weight}
\end{array}
\end{eqnarray}
We then fit the weighted $M(e^+e^-)$ distribution to obtain a luminosity-weighted number of $J/\psi$ events in each bin of $E_\gamma$ and $t$, which we denote $N^{J/\psi}_{{\mathrm{wt}}}(E_\gamma,t)$.
The energy resolution as measured by the experimental setup 
is better than the $45$~MeV bin size used in this procedure.

The cross sections are reported at the mean $t$ and $E_\gamma$ values within each bin 
(red points in Fig.~\ref{fig:dsdt_binning_data}).
Note that for a given energy region, the mean $E_\gamma$ values depend on the $t$ bin.
Still, we attribute a common mean energy within each energy region and treat
the corresponding deviations of the cross section due to the energy correction as a systematic error.
In addition, generally, the cross section averaged over the bin deviates 
from the cross section at the mean $E_\gamma$ and $t$ where it is reported,
especially for the bins that are wide and have non-rectangular shapes.
This deviation will also be treated as a systematic error.

To calculate the differential cross section, we divide the 
luminosity-weighted number of $J/\psi$ events in each bin
by the area of the bin, $a(E_\gamma,t)$,
and correct for the reconstruction efficiency $\varepsilon(E_\gamma,t)$:
\begin{eqnarray}
\begin{array}{l}
\frac{d\sigma}{dt}(E_\gamma,t) = 
\frac{N^{J/\psi}_{{\mathrm{wt}}}(E_\gamma,t)~[{\mathrm{GeV \cdot nb}}]}{a(E_\gamma,t)~[{\mathrm{GeV}}\cdot {\mathrm{GeV}}^2]}\frac{1}{\varepsilon(E_\gamma,t)}.
\label{eq:dsdt}
\end{array}
\end{eqnarray}
Thus, the differential cross section will be in units of [nb/GeV$^2$].
The area of each bin is calculated with MC by generating a uniform
distribution over the whole rectangular ($E_\gamma ,t$) plane in Fig.~\ref{fig:dsdt_binning_data}.

\begin{figure}[h]
\includegraphics[width=0.45\textwidth]{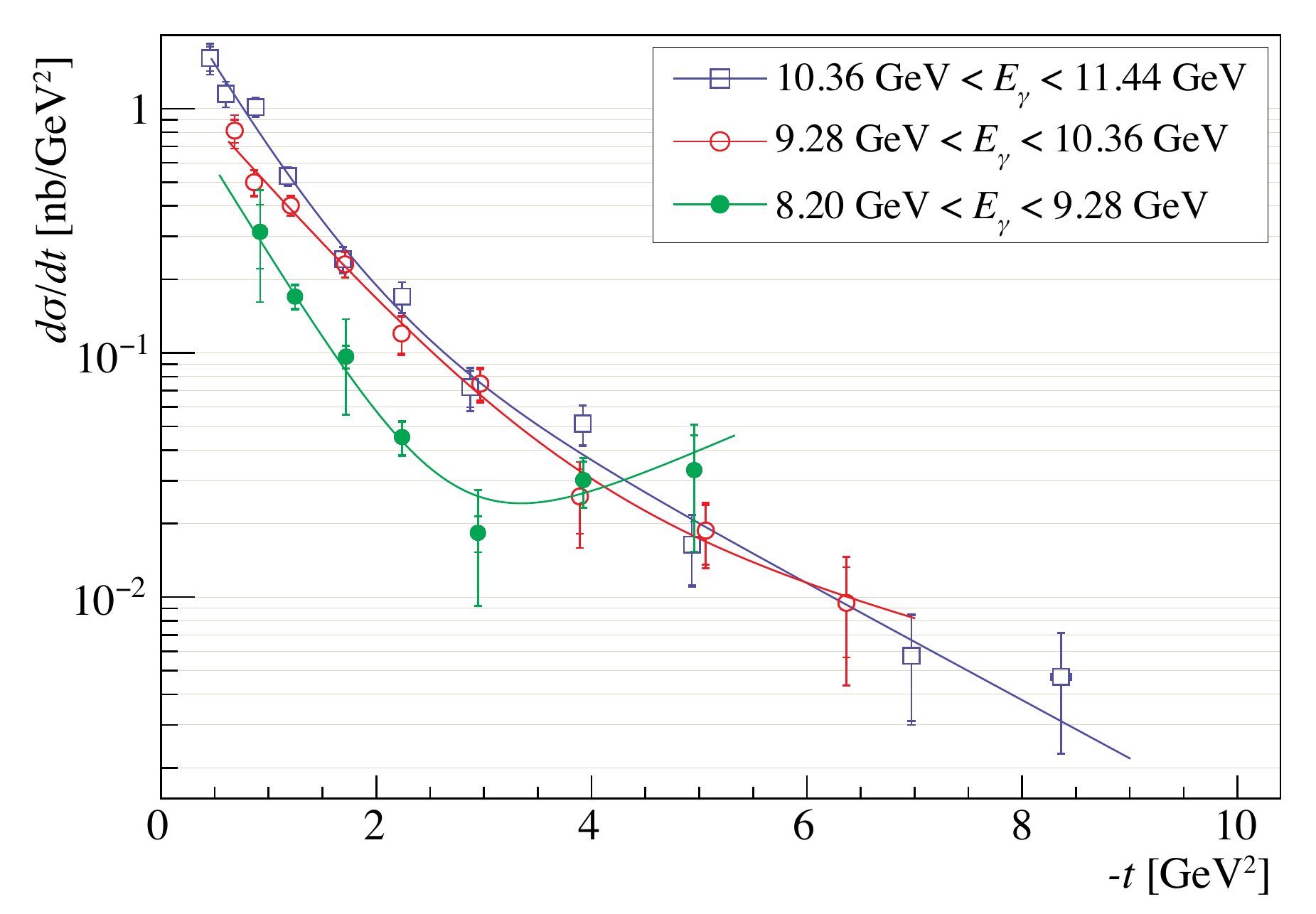}
  \caption{
The measured differential cross sections with both statistical (inner bars) and total (outer bars) uncertainties shown
for the three energy regions,
from Eq.(\ref{eq:dsdt}).
The points are fitted with a sum of two exponential functions.  
The second exponential contribution is most significant in the lowest energy
bin, where the slope changes sign.
}
  \label{fig:dsdt_iter1}
\end{figure}

We apply the same procedure for the extraction of the $J/\psi $ yields
as explained in Sec.~\ref{sec:JP} for the total cross section.
The efficiencies calculated from MC, $\varepsilon_\mathrm{MC}(E_\gamma,t)$, are corrected by the overall
normalization correction as obtained in Sec.~\ref{sec:BH}, 
using the BH process.
Thus, in Eq.(\ref{eq:dsdt}) we use
$\varepsilon(E_\gamma,t) = \varepsilon_\mathrm{MC}(E_\gamma,t)\times(0.847\pm0.019)$. 
Now we have all the ingredients in Eq.(\ref{eq:dsdt}) to calculate the
differential cross sections, and the results are given in Fig.~\ref{fig:dsdt_iter1}.
To parametrize them, they are fitted with a sum of two exponential functions.
To check the consistency of the differential cross sections, we integrate the
fitted function over the corresponding range $t_\mathrm{min}(E_{\gamma i})-t_\mathrm{max}(E_{\gamma i})$,
where $E_{\gamma i}$ is the mean energy for the corresponding energy region,
and compare these integrals with the total cross section results.
We find a good agreement, shown in Fig.~\ref{fig:sigma_dsdt_comp}.

We consider three sources in the systematic uncertainties of the individual differential data points:
(i) the uncertainty in the fitting procedure,
(ii) the correction due to the alignment of the results to a common mean energy,
and (iii) the bin-averaging effect.
To estimate the last two effects, 
we create a two-dimensional cross section model based on our measurements.
For that we use the fits of the differential cross sections in Fig.~\ref{fig:dsdt_iter1}. 
The total cross section is also fitted with a polynomial.
We note that these cross section parametrizations were used in the $J/\psi $ generator
for all the MC results presented in this paper. 
The main contribution to the systematic uncertainties for the individual data points
comes from the $J/\psi $ fitting procedure where we compare the yields extracted
from a fit with either fixed widths (based on MC) or as a free parameter, 
in the same way as was done
for the estimation of the systematic uncertainties in the total cross section.

The overall normalization uncertainty of the differential cross sections is the same
as for the total cross section, see Table \ref{tab:total_syst}. 

The numerical results for the differential cross section, along with statistical and systematic 
errors, are given in Tables~\ref{tab:dsdt0}, \ref{tab:dsdt1}, and \ref{tab:dsdt2} of the Appendix, Sec.~\ref{sec:num_res}. 
Note that in all the plots in the next section, the error bars of the GlueX data points include 
both the statistical and sytematic errors added in quadrature.

\section{Discussion}

In our cross section measurements, we observe two apparent deviations from the expectations: (i) 
of a smooth variation of the total cross section as a function of beam energy,
and (ii) of an exponentially-decreasing $t$-dependence in the differential cross sections. 
We previously mentioned the structure in the $8.8-9.4$~GeV region (Fig.~\ref{fig:sigma_dsdt_comp}) in Sec.~\ref{sec:jp_fits}.
If we treat the two points there as a potential dip, the probability that 
they are not a statistical fluctuation from a smooth fit to the observed cross sections
corresponds to a significance of $2.6\sigma$. 
However, if we consider the probability for any two adjacent points in the whole
energy interval ($8.2-11.44$~GeV) to have a deviation of at least this size,
the significance reduces to $1.4\sigma$.
Another feature that we observe is the enhancement of the
differential cross section for the lowest energy region 
towards $|t|_\mathrm{max}$ (Fig.~\ref{fig:dsdt_iter1}), which can be interpreted as an
$s$- or $u$-channel contribution.
We estimate a $2.3\sigma$ significance of such a deviation when compared to a dipole fit
of the differential cross section.
All the above significance estimates include both statistical and systematic errors.
The relevance of these features to the reaction mechanism will be discussed below.

\begin{figure}
   \includegraphics[width=0.45\textwidth]{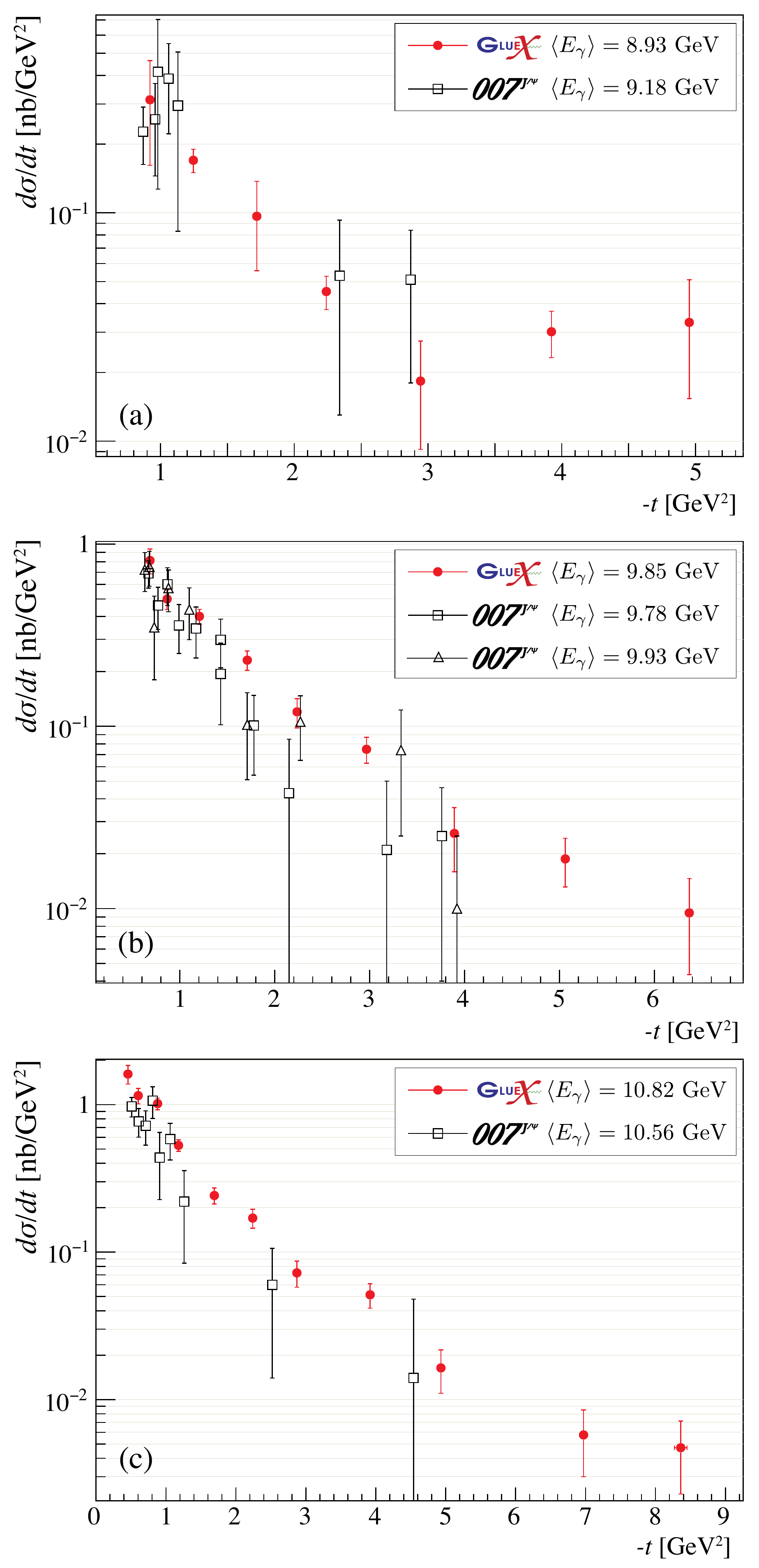}
  \caption{
Comparison of the differential cross sections for the three energy regions from this work
to the measurements of the $J/\psi-007$ experiment closest in energy~\cite{hallc_jp007}.
}\label{fig:hallc}
\end{figure}

Recently the $J/\psi-007$ experiment located in Hall C at Jefferson Lab published results on $J/\psi $
photoproduction~\cite{hallc_jp007}.
They reported $d\sigma/dt$ in 10 fine energy bins with similar
total statistics as the results reported in this paper, though in a more narrow kinematic region both in energy
and $t$.  
In Fig.~\ref{fig:hallc} we compare the GlueX results for the three energy regions
with the closest in energy differential cross sections of Ref.~\cite{hallc_jp007}.
We see good agreement between the two experiments. When comparing the two results, recall the $20\%$ scale
uncertainty in the GlueX results and note the differences in the average energies.

\begin{figure}[h]
\includegraphics[width=0.45\textwidth]{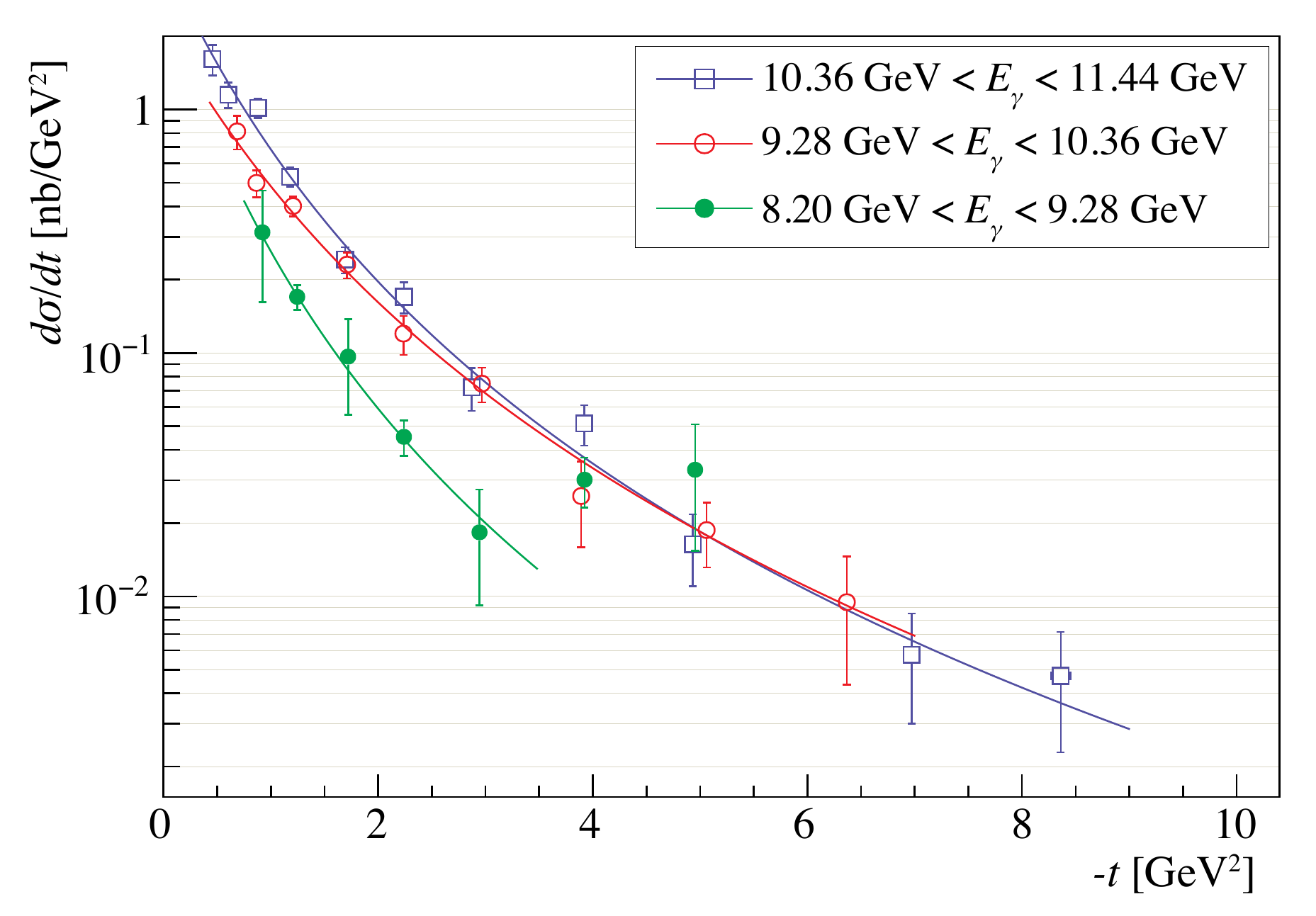}
  \caption{
The differential cross sections for the three energy regions
fitted with $[d\sigma/dt(0)]/(1-t/m_s^2)^4$, where
the cross section at $t=0$, $d\sigma/dt(0)$, and the mass scale, $m_s$, are free parameters.
}
  \label{fig:dsdt_dipole_comp}
\end{figure}
The proximity of the GlueX data to the $J/\psi$ threshold allows us to extrapolate the differential
cross sections both in beam energy and $t$ outside of the physical region 
and estimate the forward cross section at threshold, $d\sigma/dt(0)|_\mathrm{thr}$.
The forward cross section close to threshold, $d\sigma/dt(t=0, E_\gamma)$, enters in many theoretical models and plays an important role
in understanding the $J/\psi $ photoproduction and the $J/\psi $-proton 
interaction~\cite{Kharzeev99, Vanderhaeghen16, SL_1, SL_2}.
The $t$ dependence of the differential cross section can be related 
to the gluonic form factor $F(t)$ of the proton, which is usually parametrized with a dipole function, $\propto1/(1-t/m_s^2)^2$~\cite{strikman, Zahed1, Feng1, Shanahan}.
In Fig.~\ref{fig:dsdt_dipole_comp} we show the results of fits to the measured differential cross sections 
with squared dipole functions of the form $[d\sigma/dt(0)]/(1-t/m_s^2)^4$,
excluding the high-$t$ region in the lowest energy region.
The results of the fits are summarized in Table~\ref{tab:dsdt_fits}.
\begin{table}[h]
\caption{
The forward differential cross sections, $d\sigma/dt(0)$,
 and the mass scale parameter, $m_s$, from the fits shown
in Fig.~\ref{fig:dsdt_dipole_comp} for the three average beam energies, $\left<E_\gamma \right>$.  The average momentum of the final state particles in the overall center-of-mass frame, $q$, for each beam energy bin is also given.
Note, there is an overall $19.5\%$ scale uncertainty of the results for $d\sigma/dt(0)$. 
\label{tab:dsdt_fits}
}\begin{ruledtabular}
\begin{tabular}{lccc}
$\left<E_\gamma \right>$ [GeV] & $8.93$ & $9.86$ & $10.82$ \\
\colrule
$q$ [GeV] & $0.499$ & $0.767$ & $0.978$ \\
\colrule
$d\sigma/dt(0)$ [nb/GeV$^2$] & $2.863$ & $2.205$ & $4.268$ \\
& $\pm1.95$ & $\pm0.380$ & $\pm0.564$ \\
\colrule
$m_s$ [GeV] & $1.105$ & $1.472$ & $1.313$ \\
 & $\pm0.168$ & $\pm0.075$ & $\pm0.049$ \\
\end{tabular}
\end{ruledtabular}

\end{table}
The $t$-slope is defined by the mass scale parameter, $m_s$, and the fit results for $m_s$ are generally in good
agreement with the lattice calculations~\cite{Shanahan} of the $A_g(t)$ gluon form factor
that find $m_s=1.13\pm 0.06$~GeV.
More precisely, such agreement of the $J/\psi-007$ data 
(also in agreement with our data, Fig.~\ref{fig:hallc})
with the lattice calculations was demonstrated in Ref.~\cite{hallc_jp007}
using the holographic model of Ref.~\cite{Zahed2}.

\begin{figure}[t]
\includegraphics[width=0.45\textwidth]{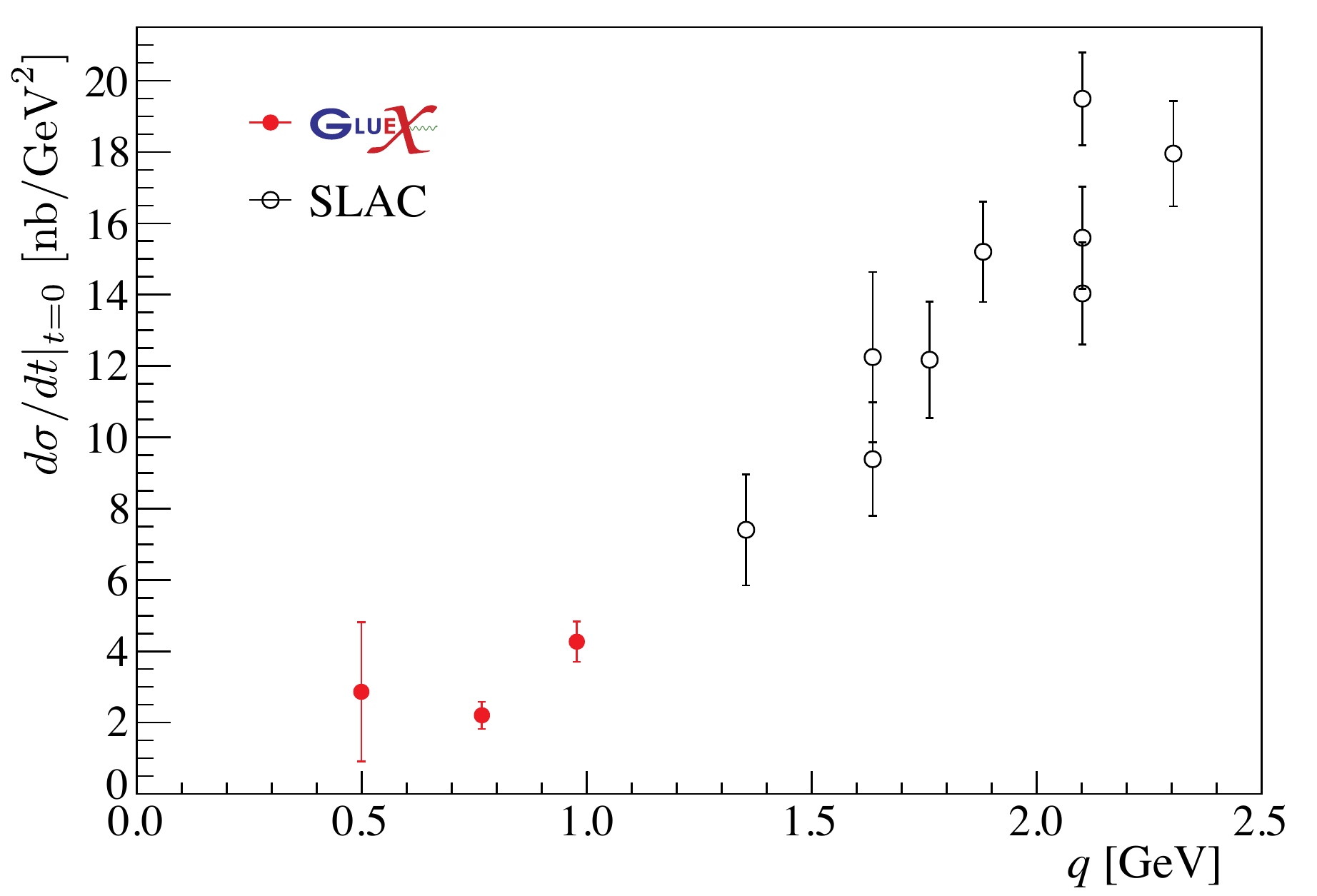}
  \caption{
The forward ($t=0$) differential cross section as a function of final particle
center-of-mass momentum from this work (filled red points) and SLAC~\cite{SLAC} measurements (open black points).
}
  \label{fig:dsdt_extra_comp}
\end{figure}

The fits in Fig.~\ref{fig:dsdt_dipole_comp} 
also directly give an extrapolation of the cross sections to $t=0$, $d\sigma/dt(0)$,
Table~\ref{tab:dsdt_fits}.
These results are plotted in
Fig.~\ref{fig:dsdt_extra_comp} as a function of the 
final proton (or $J/\psi $) c.m. momentum, $q$,
together with the SLAC measurements 
of $d\sigma/dt$ at $t=t_\mathrm{min}$
also extrapolated to $t=0$ using their measured exponential slope of $2.9$~GeV$^{-2}$~\cite{SLAC}.
Such a plot allows extrapolation of $d\sigma/dt(0)$ to the threshold, 
$d\sigma/dt(0)|_{thr.}$, that corresponds to $q=0$.
Ref.~\cite{Vanderhaeghen16} uses the VMD model and 
dispersion relations to parametrize the forward $J/\psi - p$ scattering amplitude, $T^{\psi p}$,
and to fit all existing $J/\psi$ photoproduction data
including those data taken at large center-of-mass energies.
The parametrization is then used to fit the forward differential cross sections
and estimate $d\sigma/dt(0)|_\mathrm{thr.}$ - see Fig.~3 in Ref.~\cite{Vanderhaeghen16},
which is an analog to our Fig.~\ref{fig:dsdt_extra_comp}. 
Alternatively, the extrapolation to threshold can be done by 
expanding $T^{\psi p}$ in partial waves, with the S-wave
being dominant near threshold.
Initial extrapolations were previously reported along with the preliminary GlueX results~\cite{QNP2022}, but will not be discussed further in this paper. 
It is of importance that the GlueX measurements are much closer to the threshold
than the SLAC measurements~\cite{SLAC} (the latter used in Ref.~\cite{Vanderhaeghen16}),
at the same time constraining $d\sigma/dt(0)|_\mathrm{thr.}$ to lower values
than the SLAC results and Ref.~\cite{Vanderhaeghen16}.
For the purpose of providing a quantitative estimate, let us assume $d\sigma/dt(0)|_\mathrm{thr.}$
is close in value and uncertainty to the lowest-$q$ data point in Fig.~\ref{fig:dsdt_extra_comp}, 
$2.86 \pm 2.03$~nb/GeV$^2$, where we have included the overall scale uncertainty. 
This value corresponds 
to a very small $J/\psi -p$ scattering length, $\alpha_{J/\psi p}$, which is 
given by~\cite{SL_2}:
\begin{eqnarray}
|\alpha_{J/\psi p}| =\sqrt{\frac{d\sigma}{dt}(0)\big|_\mathrm{thr.}
\frac{\gamma_\psi ^2}{\alpha \pi }\frac{k_{\gamma p}^2}{\pi }},
\label{eq:sl}
\end{eqnarray}
where $k_{\gamma p}$ is the c.m. momenta of the initial particles and 
$\gamma_\psi$ is the photon-$J/\psi $ coupling constant obtained from the
$J/\psi \to e^+e^-$ decay width.
We find $|\alpha_{J/\psi p}| = (21.3 \pm 8.2)$~$\times 10^{-3}$~fm, 
which, compared to the size of the proton of $\sim 1$~fm scale,
indicates a very weak $J/\psi -p$ interaction.
However, note that the VMD model is used in Eq.(\ref{eq:sl}) to extract this value.

We can use the mass scale $m_s$ from the fits in Fig.~\ref{fig:dsdt_dipole_comp} (Table~\ref{tab:dsdt_fits}) 
to estimate the proton mass radius as prescribed in Ref.~\cite{Kharzeev21},
\begin{eqnarray}
\sqrt{\left<r_m^2\right>} = \sqrt{\frac{6}{m_p}\frac{dG(t)}{dt}\Big|_{t=0}}=\sqrt{\frac{12}{m_s^2}},
\label{eq:pmass}
\end{eqnarray}
where the scalar gravitational form factor, $G(t)$, is related to the measured $t$-distributions
through the VMD model.
Eq.(\ref{eq:pmass}) gives $\sqrt{\left<r_m^2\right>}=$ $0.619 \pm 0.094$~fm, $0.464 \pm 0.024$~fm, and $0.521 \pm 0.020$~fm for 
$E_\gamma = 8.93$, $9.86$, and $10.82$~GeV, respectively. 
More sophisticated estimations of the proton mass radius require knowledge
of the $A(t)$ and $C(t)$ gravitational form factors separately~\cite{Ji2021},~\cite{hallc_jp007}.

\begin{figure}[]
\includegraphics[width=0.45\textwidth]{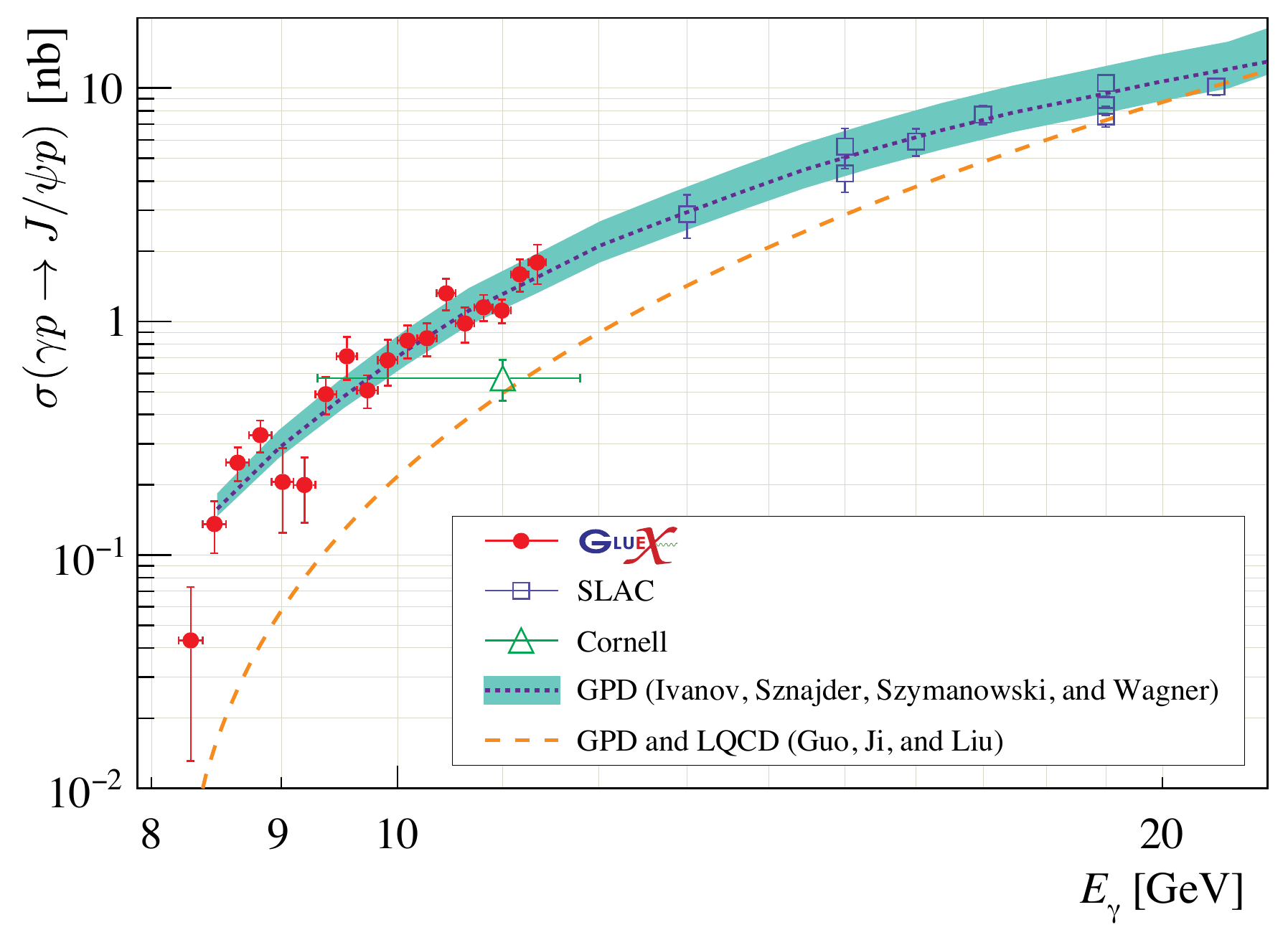}
  \caption{
Comparison of the $J/\psi $ total cross sections from this work (GlueX) to the SLAC~\cite{SLAC} and Cornell~\cite{Cornell} data 
and two QCD theoretical calculations
in the two-gluon exchange factorization model (in LO) from Ref.~\cite{Lech}
and from Ref.~\cite{Ji2021}. The latter calculation uses gravitational form factors 
from lattice calculations~\cite{Shanahan}. 
The SLAC total cross sections are estimated from their $d\sigma/dt|_{t=t_{min}}$  
measurements~\cite{SLAC} assuming a dipole $t$-dependence from the fit 
of our differential cross section at the highest energy, Fig.~\ref{fig:dsdt_dipole_comp}.
The error bars shown for the GlueX data are the statistical and systematic errors summed 
in quadrature.
} 
  \label{fig:ji1}
\end{figure}

In Fig.~\ref{fig:ji1} we compare our total cross section results to models 
that assume factorization of the  
$J/\psi$ photoproduction into a hard quark-gluon interaction 
and the GPDs describing the partonic distributions of the proton.
This factorization in exclusive heavy-meson photoproduction in terms of GPDs 
 was studied in the kinematic region of
low $|t|$ and high beam energies~\cite{Ivanov}.
The factorization was explicitly demonstrated
 by direct leading order (LO) and next-to-leading order (NLO) calculations.
In Ref.~\cite{Ji2021}, it was shown that in the limit of high meson masses and at LO, 
the factorization in terms of gluon GPDs is still valid down to the threshold. 
Calculations in this framework were performed for the $J/\psi $ photoproduction cross section
using parametrizations of the gravitational form factors
obtained from the lattice results of Ref.~\cite{Shanahan}.
These calculations for the total cross section are compared to our measurements in Fig.~\ref{fig:ji1}.
While they agree better with the SLAC data at higher energies, 
they underestimate our near-threshold measurements.
Recently, the authors of Ref.~\cite{Ivanov}
extended their calculations to the threshold region at LO~\cite{Lech}.
These calculations, plotted also in Fig.~\ref{fig:ji1},
are in a very good agreement with the total cross section measurements. 
Attempts to include the NLO contribution result in large uncertainties
due to the poor knowledge of the corresponding GPD functions in this kinematic region~\cite{Jakub}.
This indicates that our measurements can strongly constrain the relevant gluon GPD functions.

\begin{figure}[]
\includegraphics[width=0.45\textwidth]{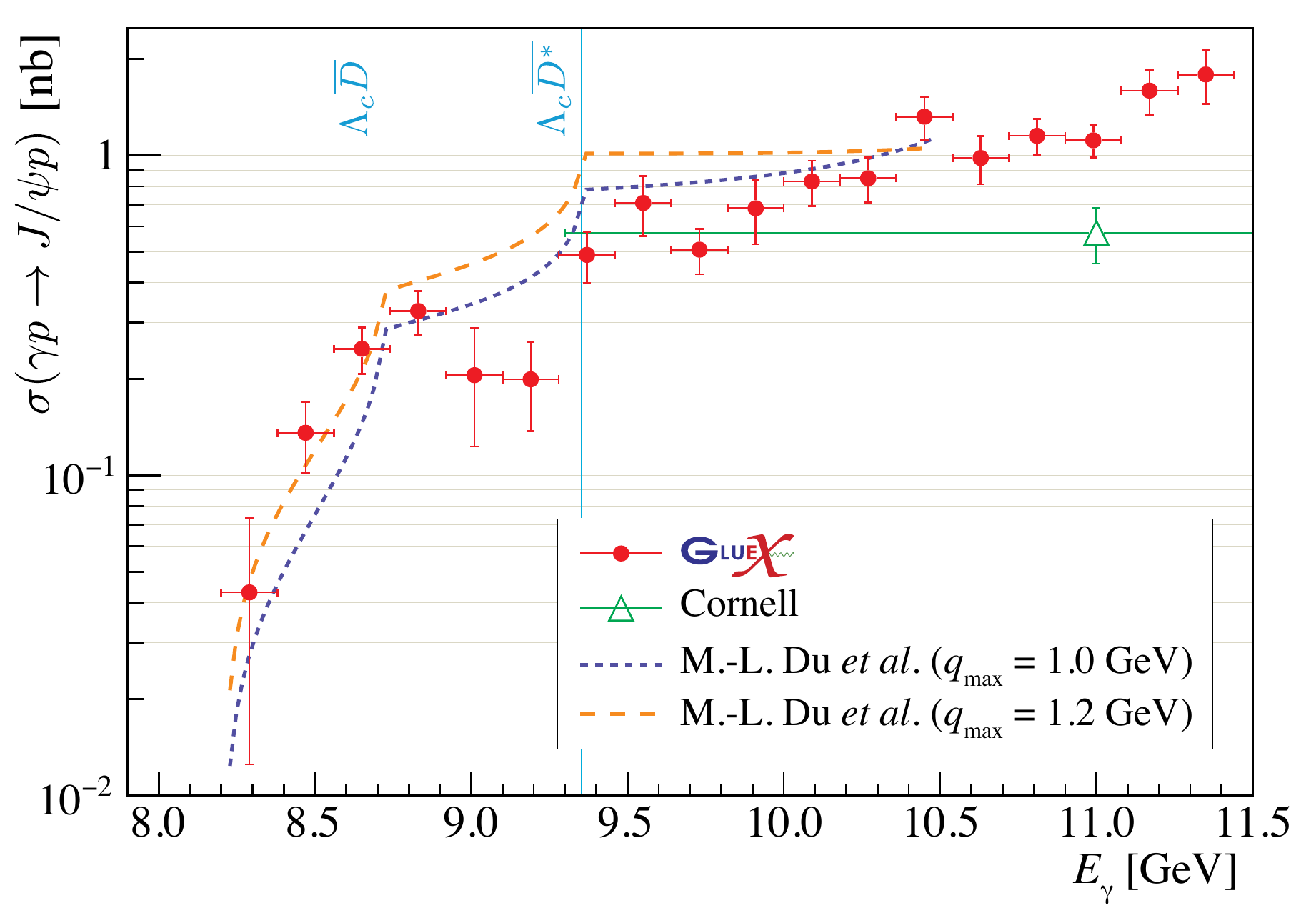}
  \caption{
Comparison of the GlueX $J/\psi $ total cross section to open charm 
calculations~\cite{openc}.
The thresholds of $\Lambda_c\bar{D}$ ($8.71$~GeV) 
and $\Lambda_c\bar{D}^*$ ($9.35$~GeV) 
are shown as vertical lines. 
The error bars shown for the GlueX data are the statistical and systematic errors summed 
in quadrature.
} 
  \label{fig:openc}
\end{figure}
The authors of Ref.~\cite{openc} propose an alternative 
mechanism of $J/\psi $ photoproduction 
with a dominant exchange of open-charm channels $\Lambda_c\bar{D}$ and $\Lambda_c\bar{D}^*$ in box diagrams.
We show the total cross section results of this model in Fig.~\ref{fig:openc}, and find good qualitative agreement with our measurements.
In particular, in the data we see structures peaking at both the $\Lambda_c\bar{D}$ and $\Lambda_c\bar{D}^*$ 
thresholds that can be interpreted as the cusps expected with this reaction mechanism.
However, the exchange of heavy hadrons in this model implies 
a very shallow $t$-dependence in the differential cross sections.
This is not supported by the steeply falling cross sections we observe, as shown in Fig.~\ref{fig:dsdt_dipole_comp}. 
Therefore, our differential cross section measurements do not support a dominant contribution from these open charm exchanges, although the enhancement at high $t$ observed for the lowest beam energy region is consistent with a possible contribution from these exchanges.
Alternatively, in Ref.~\cite{Lech_u_channel} it was shown that the high-$t$ enhancement 
can be explained by $u$-channel contribution assuming factorization in terms of 
Transition Distribution Amplitudes~\cite{TDA}.

In Ref.~\cite{JPAC_paper}, the model-independent effective range expansion was used to parameterize the lowest partial waves.
Fits  to the total and differential cross sections from this paper and from Ref.~\cite{hallc_jp007} show that the expansion is rapidly
 convergent, with the $L\le 3$ waves saturating the forward peak in the measured photon energy range.
Furthermore, the energy dependence of the
  total cross section near the open-charm thresholds 
 was shown to be consistent with the appearance of
 $\Lambda_c \bar{D}^{(*)}$ intermediate states,
 as suggested by Ref.~\cite{openc}.

It is important to be able to understand the dynamics underlying
$J/\psi$ photoproduction at threshold, and possibly to identify a
kinematic region that can be used to extract the proton gluonic form
factors. Based on the $t$-slopes of the differential cross sections
(Fig.~\ref{fig:dsdt_dipole_comp}) and also the results of
Ref.~\cite{hallc_jp007}, the differential cross section at low
$t$-values is consistent with being dominantly due to gluonic
exchange.  
However, the possible structures in the total cross section energy
dependence and the flattening of the differential cross section near
threshold are consistent with contributions from open-charm
intermediate states.
So far, from the analyses of Ref.~\cite{JPAC_paper} it is not possible
to distinguish
between the gluon and open-charm exchange mechanisms.
Certainly, further theoretical work is needed to understand the
mechanism of near-threshold $J/\psi$ production and its relation to
the gluonic structure of the proton, especially since hints of open
charm production are visible. On the experimental side, higher
statistics are needed to confirm the structures in the total cross
section and the enhancement in the $t$-dependence, the statistical
significance of which at present does not allow making of definitive
conclusions.

\vspace*{1.cm}

\begin{acknowledgments}
We would like to thank A. N. H. Blin, A. Pilloni, A. P. Szczepaniak, and
D. Winney for the fruitful discussions of the interpretation of the results.  
We would like to acknowledge the outstanding efforts of the staff of the Accelerator and the Physics Divisions at Jefferson Lab that made the experiment possible. This work was supported in part by the U.S. Department of Energy, the U.S. National Science Foundation, the German Research Foundation, Forschungszentrum J\"ulich GmbH, GSI Helmholtzzentrum f\"ur Schwerionenforschung GmbH, the Natural Sciences and Engineering Research Council of Canada, the Russian Foundation for Basic Research, the UK Science and Technology Facilities Council, the Chilean Comisi\'{o}n Nacional de Investigaci\'{o}n Cient\'{i}fica y Tecnol\'{o}gica, the National Natural Science Foundation of China and the China Scholarship Council. This material is based upon work supported by the U.S. Department of Energy, Office of Science, Office of Nuclear Physics under contract DE-AC0506OR23177.
This research used resources of the National Energy Research Scientific Computing Center (NERSC), a U.S. Department of Energy Office of Science User Facility operated under Contract No. DE-AC02-05CH11231. This work used the Extreme Science and Engineering Discovery Environment (XSEDE), which is supported by National Science Foundation grant number ACI-1548562. Specifically, it used the Bridges system, which is supported by NSF award number ACI-1445606, at the Pittsburgh Supercomputing Center (PSC).
\end{acknowledgments}

\appendix

\section{Numerical results}
\label{sec:num_res}

\begin{table}[!h]
\caption{$\gamma p \rightarrow J/\psi p$ total cross sections in bins of beam energy. 
 The first uncertainties are statistical, and the second are systematic. }
\begin{ruledtabular}
\begin{tabular}{D{-}{-}{-1}c}
\multicolumn{1}{c}{Energy bin [GeV]}&
\textrm{$\sigma$ [nb]} \\
\colrule
 8.20-8.38 & $0.043\pm0.012\pm0.027$ \\
 8.38-8.56 & $0.136\pm0.022\pm0.026$ \\
 8.56-8.74 & $0.249\pm0.029\pm0.029$\\
 8.74-8.92 & $0.326\pm0.048\pm0.016$\\
 8.92-9.10 & $0.206\pm0.059\pm0.056$\\
 9.10-9.28 & $0.200\pm0.060\pm0.018$\\
 9.28-9.46 & $0.489\pm0.087\pm0.019$\\
 9.46-9.64 & $0.710\pm0.134\pm0.064$\\
 9.64-9.82 & $0.507\pm0.080\pm0.019$\\
 9.82-10.00 & $0.683\pm0.100\pm0.116$\\
 10.00-10.18 & $0.829\pm0.119\pm0.064$\\
 10.18-10.36 & $0.848\pm0.123\pm0.059$\\
 10.36-10.54 & $1.321\pm0.193\pm0.067$\\
 10.54-10.72& $0.981\pm0.134\pm0.104$\\
 10.72-10.90 & $1.151\pm0.140\pm0.051$\\
 10.90-11.08 & $1.114\pm0.126\pm0.034$\\
 11.08-11.26 & $1.594\pm0.208\pm0.144$\\
 11.26-11.44 & $1.791\pm0.344\pm0.026$\\

\end{tabular}
\end{ruledtabular}
\label{tab:sigma}
\end{table}

\begin{table}[!h]
\caption{$\gamma p \rightarrow J/\psi p$ differential cross sections
in the $8.2-9.28$~GeV beam energy range, 
average $t$ and beam energy in bins of $t$. 
The first cross section uncertainties are statistical, and the second are systematic.
The overall average beam energy is $8.93$~GeV.} 
\begin{ruledtabular}
\begin{tabular}{lccc}
\textrm{$t$ bin}&
\textrm{$\left<t\right>$ }&
\textrm{$\left<E_\gamma\right>$}&
\textrm{$d\sigma/dt$ }\\
{[GeV$^2$]} &  {[GeV$^2$]} & [GeV] & \multicolumn{1}{c}{$\mathrm{[nb}/\mathrm{GeV}^{2}]$} \\
\colrule
 $0.77-1.00$&0.92& 9.14 & $0.313\pm0.092\pm0.120$\\
 $1.00-1.50$&1.25& 8.96 & $0.170\pm0.018\pm0.008$\\
 $1.50-2.00$&1.72& 8.80 & $0.097\pm0.010\pm0.040$\\
 $2.00-2.50$&2.24& 8.77 & $0.045\pm0.007\pm0.003$\\
 $2.50-3.50$&2.94& 8.78 & $0.018\pm0.003\pm0.009$\\
 $3.50-4.50$&3.92& 8.95 & $0.030\pm0.006\pm0.004$\\
 $4.50-5.75$&4.95& 9.10 & $0.033\pm0.013\pm0.012$\\
\end{tabular}
\end{ruledtabular}

\label{tab:dsdt0}
\end{table}

\begin{table}[!h]
\caption{$\gamma p \rightarrow J/\psi p$ differential cross sections 
in the $9.28-10.36$~GeV beam energy range, 
average $t$ and beam energy in bins of $t$. 
 The first cross section uncertainties are statistical, and the second are systematic.
The overall average beam energy is $9.86$~GeV.}
\begin{ruledtabular}
\begin{tabular}{lccc}
\textrm{$t$ bin}&
\textrm{$\left<t\right>$}&
\textrm{$\left<E_\gamma\right>$ }&
\textrm{$d\sigma/dt$ }\\
{[GeV$^2$]} &  {[GeV$^2$]} & [GeV] & \multicolumn{1}{c}{$\mathrm{[nb}/\mathrm{GeV}^{2}]$} \\
\colrule
 $0.49-0.77$&0.69& 10.00 & $0.813\pm0.088\pm0.092$\\
 $0.77-1.00$&0.87& 9.85 & $0.499\pm0.061\pm0.016$\\
 $1.00-1.50$&1.21& 9.83 & $0.401\pm0.037\pm0.010$\\
 $1.50-2.00$&1.71& 9.83 & $0.231\pm0.027\pm0.006$\\
 $2.00-2.50$&2.24& 9.82 & $0.120\pm0.021\pm0.007$\\
 $2.50-3.50$&2.97& 9.84 & $0.075\pm0.011\pm0.005$\\
 $3.50-4.50$&3.89& 9.86 & $0.026\pm0.008\pm0.006$\\
 $4.50-5.75$&5.06& 9.76 & $0.019\pm0.005\pm0.002$\\
 $5.75-8.10$&6.37& 9.93 & $0.009\pm0.004\pm0.003$\\
\end{tabular}
\end{ruledtabular}

\label{tab:dsdt1}
\end{table}

\begin{table}[!h]
\caption{$\gamma p \rightarrow J/\psi p$ differential cross sections
in the $10.36-11.44$~GeV beam energy range, 
average $t$ and energy, in bins of $t$. 
 The first cross section uncertainties are statistical, and the second are systematic.
The overall average beam energy is $10.82$~GeV.}
\begin{ruledtabular}
\begin{tabular}{lccc}
\textrm{$t$ bin}&
\textrm{$\left<t\right>$}&
\textrm{$\left<E_\gamma\right>$}&
\multicolumn{1}{c}{$d\sigma/dt$} \\
{[GeV$^2$]} &  {[GeV$^2$]} & [GeV] & \multicolumn{1}{c}{$\mathrm{[nb}/\mathrm{GeV}^{2}]$} \\
\colrule
 $0.35-0.49$&0.46& 10.96 & $1.611\pm0.187\pm0.139$\\
 $0.49-0.77$&0.60& 10.87 & $1.150\pm0.084\pm0.109$\\
 $0.77-1.00$&0.88& 10.85 & $1.015\pm0.089\pm0.023$\\
 $1.00-1.50$&1.18& 10.86 & $0.529\pm0.042\pm0.023$\\
 $1.50-2.00$&1.69& 10.86 & $0.242\pm0.029\pm0.008$\\
 $2.00-2.50$&2.24& 10.83 & $0.170\pm0.025\pm0.003$\\
 $2.50-3.50$&2.87& 10.82 & $0.072\pm0.012\pm0.008$\\
 $3.50-4.50$&3.92& 10.81 & $0.051\pm0.009\pm0.002$\\
 $4.50-5.75$&4.93& 10.78 & $0.016\pm0.005\pm0.001$\\
 $5.75-8.10$&6.97& 10.70 & $0.0058\pm0.0026\pm0.0008$\\
 $8.10-10.30$&8.36& 10.70 & $0.0047\pm0.0024\pm0.0002$\\
\end{tabular}
\end{ruledtabular}

\label{tab:dsdt2}
\end{table}

\bibliography{jpsi_prc.bib}

\end{document}

%% file: authors.tex
\affiliation{Arizona State University, Tempe, Arizona 85287, USA}
\affiliation{National and Kapodistrian University of Athens, 15771 Athens, Greece}
\affiliation{Helmholtz-Institut f\"{u}r Strahlen- und Kernphysik Universit\"{a}t Bonn, D-53115 Bonn, Germany}
\affiliation{Carnegie Mellon University, Pittsburgh, Pennsylvania 15213, USA}
\affiliation{The Catholic University of America, Washington, D.C. 20064, USA}
\affiliation{University of Connecticut, Storrs, Connecticut 06269, USA}
\affiliation{Duke University, Durham, North Carolina 27708, USA}
\affiliation{Florida International University, Miami, Florida 33199, USA}
\affiliation{Florida State University, Tallahassee, Florida 32306, USA}
\affiliation{The George Washington University, Washington, D.C. 20052, USA}
\affiliation{University of Glasgow, Glasgow G12 8QQ, United Kingdom}
\affiliation{GSI Helmholtzzentrum f\"{u}r Schwerionenforschung GmbH, D-64291 Darmstadt, Germany}
\affiliation{Institute of High Energy Physics, Beijing 100049, People's Republic of China}
\affiliation{Indiana University, Bloomington, Indiana 47405, USA}
\affiliation{IKP, Forschungszentrum J\"{u}lich, D-52428 J\"{u}lich GmbH, Germany}
\affiliation{National Research Centre Kurchatov Institute, Moscow 123182, Russia}
\affiliation{Lamar University, Beaumont, Texas 77710, USA}
\affiliation{Johannes Gutenberg-Universit\"{a}t Mainz, 55128 Mainz, Germany}
\affiliation{University of Massachusetts, Amherst, Massachusetts 01003, USA}
\affiliation{Massachusetts Institute of Technology, Cambridge, Massachusetts 02139, USA}
\affiliation{National Research Nuclear University Moscow Engineering Physics Institute, Moscow 115409, Russia}
\affiliation{Mount Allison University, Sackville, New Brunswick E4L 1E6, Canada}
\affiliation{Norfolk State University, Norfolk, Virginia 23504, USA}
\affiliation{North Carolina A\&T State University, Greensboro, North Carolina 27411, USA}
\affiliation{University of North Carolina at Wilmington, Wilmington, North Carolina 28403, USA}
\affiliation{Old Dominion University, Norfolk, Virginia 23529, USA}
\affiliation{University of Regina, Regina, Saskatchewan S4S 0A2, Canada}
\affiliation{Universidad T\'ecnica Federico Santa Mar\'ia, Casilla 110-V Valpara\'iso, Chile}
\affiliation{Thomas Jefferson National Accelerator Facility, Newport News, Virginia 23606, USA}
\affiliation{Tomsk Polytechnic University, 634050 Tomsk, Russia}
\affiliation{Tomsk State University, 634050 Tomsk, Russia}
\affiliation{Union College, Schenectady, New York 12308, USA}
\affiliation{Washington \& Jefferson College, Washington, Pennsylvania 15301, USA}
\affiliation{William \& Mary, Williamsburg, Virginia 23185, USA}
\affiliation{Wuhan University, Wuhan, Hubei 430072, People's Republic of China}
\affiliation{A. I. Alikhanian National Science Laboratory (Yerevan Physics Institute), 0036 Yerevan, Armenia}
\author{S.~Adhikari} \affiliation{Old Dominion University, Norfolk, Virginia 23529, USA}
\author{F.~Afzal,\orcidlink{0000-0001-8063-6719 }} \affiliation{Helmholtz-Institut f\"{u}r Strahlen- und Kernphysik Universit\"{a}t Bonn, D-53115 Bonn, Germany}
\author{C.~S.~Akondi,\orcidlink{0000-0001-6303-5217}} \affiliation{Florida State University, Tallahassee, Florida 32306, USA}
\author{M.~Albrecht,\orcidlink{0000-0001-6180-4297}} \affiliation{Thomas Jefferson National Accelerator Facility, Newport News, Virginia 23606, USA}
\author{M.~Amaryan} \affiliation{Old Dominion University, Norfolk, Virginia 23529, USA}
\author{V.~Arroyave} \affiliation{Florida International University, Miami, Florida 33199, USA}
\author{A.~Asaturyan,\orcidlink{0000-0002-8105-913X}} \affiliation{University of North Carolina at Wilmington, Wilmington, North Carolina 28403, USA}\affiliation{A. I. Alikhanian National Science Laboratory (Yerevan Physics Institute), 0036 Yerevan, Armenia}
\author{A.~Austregesilo,\orcidlink{0000-0002-9291-4429}} \affiliation{Thomas Jefferson National Accelerator Facility, Newport News, Virginia 23606, USA}
\author{Z.~Baldwin,\orcidlink{0000-0002-8534-0922}} \affiliation{Carnegie Mellon University, Pittsburgh, Pennsylvania 15213, USA}
\author{F.~Barbosa} \affiliation{Thomas Jefferson National Accelerator Facility, Newport News, Virginia 23606, USA}
% ...
\collaboration{The \textsc{GlueX} Collaboration}

%% file: jpsi_prc.bbl
\begin{thebibliography}{52}%
\makeatletter
\providecommand \@ifxundefined [1]{%
 \@ifx{#1\undefined}
}%
\providecommand \@ifnum [1]{%
 \ifnum #1\expandafter \@firstoftwo
 \else \expandafter \@secondoftwo
 \fi
}%
\providecommand \@ifx [1]{%
 \ifx #1\expandafter \@firstoftwo
 \else \expandafter \@secondoftwo
 \fi
}%
\providecommand \natexlab [1]{#1}%
\providecommand \enquote  [1]{``#1''}%
\providecommand \bibnamefont  [1]{#1}%
\providecommand \bibfnamefont [1]{#1}%
\providecommand \citenamefont [1]{#1}%
\providecommand \href@noop [0]{\@secondoftwo}%
\providecommand \href [0]{\begingroup \@sanitize@url \@href}%
\providecommand \@href[1]{\@@startlink{#1}\@@href}%
\providecommand \@@href[1]{\endgroup#1\@@endlink}%
\providecommand \@sanitize@url [0]{\catcode `\\12\catcode `\$12\catcode
  `\&12\catcode `\#12\catcode `\^12\catcode `\_12\catcode `\%12\relax}%
\providecommand \@@startlink[1]{}%
\providecommand \@@endlink[0]{}%
\providecommand \url  [0]{\begingroup\@sanitize@url \@url }%
\providecommand \@url [1]{\endgroup\@href {#1}{\urlprefix }}%
\providecommand \urlprefix  [0]{URL }%
\providecommand \Eprint [0]{\href }%
\providecommand \doibase [0]{http://dx.doi.org/}%
\providecommand \selectlanguage [0]{\@gobble}%
\providecommand \bibinfo  [0]{\@secondoftwo}%
\providecommand \bibfield  [0]{\@secondoftwo}%
\providecommand \translation [1]{[#1]}%
\providecommand \BibitemOpen [0]{}%
\providecommand \bibitemStop [0]{}%
\providecommand \bibitemNoStop [0]{.\EOS\space}%
\providecommand \EOS [0]{\spacefactor3000\relax}%
\providecommand \BibitemShut  [1]{\csname bibitem#1\endcsname}%
\let\auto@bib@innerbib\@empty
\bibitem [{\citenamefont {Ali~{et al. (GlueX
  collaboration)}}(2019)}]{prl_gluex}%
  \BibitemOpen
  \bibfield  {author} {\bibinfo {author} {\bibfnamefont {A.}~\bibnamefont
  {Ali~{et al. (GlueX collaboration)}}},\ }\href@noop {} {\bibfield  {journal}
  {\bibinfo  {journal} {Phys. Rev. Lett.}\ }\textbf {\bibinfo {volume} {123}},\
  \bibinfo {pages} {072001} (\bibinfo {year} {2019})}\BibitemShut {NoStop}%
\bibitem [{\citenamefont {Frankfurt}\ and\ \citenamefont
  {Strikman}(2002)}]{strikman}%
  \BibitemOpen
  \bibfield  {author} {\bibinfo {author} {\bibfnamefont {L.}~\bibnamefont
  {Frankfurt}}\ and\ \bibinfo {author} {\bibfnamefont {M.}~\bibnamefont
  {Strikman}},\ }\href@noop {} {\bibfield  {journal} {\bibinfo  {journal}
  {Phys. Rev. D}\ }\textbf {\bibinfo {volume} {66}},\ \bibinfo {pages} {031502}
  (\bibinfo {year} {2002})}\BibitemShut {NoStop}%
\bibitem [{\citenamefont {Kharzeev}\ \emph {et~al.}(1999)\citenamefont
  {Kharzeev}, \citenamefont {Satz}, \citenamefont {Syamtomov},\ and\
  \citenamefont {Zinovev}}]{Kharzeev99}%
  \BibitemOpen
  \bibfield  {author} {\bibinfo {author} {\bibfnamefont {D.}~\bibnamefont
  {Kharzeev}}, \bibinfo {author} {\bibfnamefont {H.}~\bibnamefont {Satz}},
  \bibinfo {author} {\bibfnamefont {A.}~\bibnamefont {Syamtomov}}, \ and\
  \bibinfo {author} {\bibfnamefont {G.}~\bibnamefont {Zinovev}},\ }\href@noop
  {} {\bibfield  {journal} {\bibinfo  {journal} {Nucl.Phys. A}\ }\textbf
  {\bibinfo {volume} {661}},\ \bibinfo {pages} {568} (\bibinfo {year}
  {1999})}\BibitemShut {NoStop}%
\bibitem [{\citenamefont {Hatta}\ and\ \citenamefont {Yang}(2018)}]{Hatta1}%
  \BibitemOpen
  \bibfield  {author} {\bibinfo {author} {\bibfnamefont {Y.}~\bibnamefont
  {Hatta}}\ and\ \bibinfo {author} {\bibfnamefont {D.-L.}\ \bibnamefont
  {Yang}},\ }\href {\doibase 10.1103/physrevd.98.074003} {\bibfield  {journal}
  {\bibinfo  {journal} {Phys. Rev. D}\ }\textbf {\bibinfo {volume} {98}}
  (\bibinfo {year} {2018}),\ 10.1103/physrevd.98.074003}\BibitemShut {NoStop}%
\bibitem [{\citenamefont {Kou}\ \emph {et~al.}(2021)\citenamefont {Kou},
  \citenamefont {Wang},\ and\ \citenamefont {Chen}}]{Wang2}%
  \BibitemOpen
  \bibfield  {author} {\bibinfo {author} {\bibfnamefont {W.}~\bibnamefont
  {Kou}}, \bibinfo {author} {\bibfnamefont {R.}~\bibnamefont {Wang}}, \ and\
  \bibinfo {author} {\bibfnamefont {X.}~\bibnamefont {Chen}},\ }\href@noop {}
  {\enquote {\bibinfo {title} {Determination of the gluonic d-term and
  mechanical radii of proton from experimental data},}\ } (\bibinfo {year}
  {2021}),\ \Eprint {http://arxiv.org/abs/2104.12962} {arXiv:2104.12962
  [hep-ph]} \BibitemShut {NoStop}%
\bibitem [{\citenamefont {Strakovsky}\ \emph {et~al.}(2020)\citenamefont
  {Strakovsky}, \citenamefont {Epifanov},\ and\ \citenamefont
  {Pentchev}}]{SL_1}%
  \BibitemOpen
  \bibfield  {author} {\bibinfo {author} {\bibfnamefont {I.~I.}\ \bibnamefont
  {Strakovsky}}, \bibinfo {author} {\bibfnamefont {D.}~\bibnamefont
  {Epifanov}}, \ and\ \bibinfo {author} {\bibfnamefont {L.}~\bibnamefont
  {Pentchev}},\ }\href {\doibase 10.1103/physrevc.101.042201} {\bibfield
  {journal} {\bibinfo  {journal} {Phys. Rev. C}\ }\textbf {\bibinfo {volume}
  {101}} (\bibinfo {year} {2020}),\ 10.1103/physrevc.101.042201}\BibitemShut
  {NoStop}%
\bibitem [{\citenamefont {Pentchev}\ and\ \citenamefont
  {Strakovsky}(2021)}]{SL_2}%
  \BibitemOpen
  \bibfield  {author} {\bibinfo {author} {\bibfnamefont {L.}~\bibnamefont
  {Pentchev}}\ and\ \bibinfo {author} {\bibfnamefont {I.~I.}\ \bibnamefont
  {Strakovsky}},\ }\href {\doibase 10.1140/epja/s10050-021-00364-4} {\bibfield
  {journal} {\bibinfo  {journal} {Eur.Phys.J.A}\ }\textbf {\bibinfo {volume}
  {57}} (\bibinfo {year} {2021}),\ 10.1140/epja/s10050-021-00364-4}\BibitemShut
  {NoStop}%
\bibitem [{\citenamefont {Ivanov}\ \emph {et~al.}(2004)\citenamefont {Ivanov},
  \citenamefont {Schafer}, \citenamefont {Szymanowski},\ and\ \citenamefont
  {Krasnikov}}]{Ivanov}%
  \BibitemOpen
  \bibfield  {author} {\bibinfo {author} {\bibfnamefont {D.~Y.}\ \bibnamefont
  {Ivanov}}, \bibinfo {author} {\bibfnamefont {A.}~\bibnamefont {Schafer}},
  \bibinfo {author} {\bibfnamefont {L.}~\bibnamefont {Szymanowski}}, \ and\
  \bibinfo {author} {\bibfnamefont {G.}~\bibnamefont {Krasnikov}},\ }\href
  {\doibase 10.1140/epjc/s2004-01712-x} {\bibfield  {journal} {\bibinfo
  {journal} {Eur. Phys. J. C}\ }\textbf {\bibinfo {volume} {34}},\ \bibinfo
  {pages} {297} (\bibinfo {year} {2004})},\ \bibinfo {note} {[Erratum:
  Eur.Phys.J.C 75, 75 (2015)]},\ \Eprint {http://arxiv.org/abs/hep-ph/0401131}
  {arXiv:hep-ph/0401131} \BibitemShut {NoStop}%
\bibitem [{\citenamefont {Hatta}\ and\ \citenamefont
  {Strikman}(2021)}]{Hatta_Strikman}%
  \BibitemOpen
  \bibfield  {author} {\bibinfo {author} {\bibfnamefont {Y.}~\bibnamefont
  {Hatta}}\ and\ \bibinfo {author} {\bibfnamefont {M.}~\bibnamefont
  {Strikman}},\ }\href {\doibase 10.1016/j.physletb.2021.136295} {\bibfield
  {journal} {\bibinfo  {journal} {Phys. Lett. B}\ }\textbf {\bibinfo {volume}
  {817}},\ \bibinfo {pages} {136295} (\bibinfo {year} {2021})}\BibitemShut
  {NoStop}%
\bibitem [{\citenamefont {Guo}\ \emph {et~al.}(2021)\citenamefont {Guo},
  \citenamefont {Ji},\ and\ \citenamefont {Liu}}]{Ji2021}%
  \BibitemOpen
  \bibfield  {author} {\bibinfo {author} {\bibfnamefont {Y.}~\bibnamefont
  {Guo}}, \bibinfo {author} {\bibfnamefont {X.}~\bibnamefont {Ji}}, \ and\
  \bibinfo {author} {\bibfnamefont {Y.}~\bibnamefont {Liu}},\ }\href {\doibase
  10.1103/PhysRevD.103.096010} {\bibfield  {journal} {\bibinfo  {journal}
  {Phys. Rev. D}\ }\textbf {\bibinfo {volume} {103}},\ \bibinfo {pages}
  {096010} (\bibinfo {year} {2021})},\ \Eprint
  {http://arxiv.org/abs/2103.11506} {arXiv:2103.11506 [hep-ph]} \BibitemShut
  {NoStop}%
\bibitem [{\citenamefont {Kharzeev}(2021)}]{Kharzeev21}%
  \BibitemOpen
  \bibfield  {author} {\bibinfo {author} {\bibfnamefont {D.~E.}\ \bibnamefont
  {Kharzeev}},\ }\href {\doibase 10.1103/physrevd.104.054015} {\bibfield
  {journal} {\bibinfo  {journal} {Physical Review D}\ }\textbf {\bibinfo
  {volume} {104}} (\bibinfo {year} {2021}),\
  10.1103/physrevd.104.054015}\BibitemShut {NoStop}%
\bibitem [{\citenamefont {Ji}\ \emph {et~al.}(2021)\citenamefont {Ji},
  \citenamefont {Liu},\ and\ \citenamefont {Zahed}}]{Jimass}%
  \BibitemOpen
  \bibfield  {author} {\bibinfo {author} {\bibfnamefont {X.}~\bibnamefont
  {Ji}}, \bibinfo {author} {\bibfnamefont {Y.}~\bibnamefont {Liu}}, \ and\
  \bibinfo {author} {\bibfnamefont {I.}~\bibnamefont {Zahed}},\ }\href
  {\doibase 10.1103/physrevd.103.074002} {\bibfield  {journal} {\bibinfo
  {journal} {Phys. Rev. D}\ }\textbf {\bibinfo {volume} {103}} (\bibinfo {year}
  {2021}),\ 10.1103/physrevd.103.074002}\BibitemShut {NoStop}%
\bibitem [{\citenamefont {Mamo}\ and\ \citenamefont
  {Zahed}(2021{\natexlab{a}})}]{Zahed2}%
  \BibitemOpen
  \bibfield  {author} {\bibinfo {author} {\bibfnamefont {K.~A.}\ \bibnamefont
  {Mamo}}\ and\ \bibinfo {author} {\bibfnamefont {I.}~\bibnamefont {Zahed}},\
  }\href {\doibase 10.1103/physrevd.103.094010} {\bibfield  {journal} {\bibinfo
   {journal} {Phys. Rev. D}\ }\textbf {\bibinfo {volume} {103}} (\bibinfo
  {year} {2021}{\natexlab{a}}),\ 10.1103/physrevd.103.094010}\BibitemShut
  {NoStop}%
\bibitem [{\citenamefont {Wang}\ \emph {et~al.}(2021)\citenamefont {Wang},
  \citenamefont {Kou}, \citenamefont {Xie},\ and\ \citenamefont
  {Chen}}]{Wang1}%
  \BibitemOpen
  \bibfield  {author} {\bibinfo {author} {\bibfnamefont {R.}~\bibnamefont
  {Wang}}, \bibinfo {author} {\bibfnamefont {W.}~\bibnamefont {Kou}}, \bibinfo
  {author} {\bibfnamefont {Y.-P.}\ \bibnamefont {Xie}}, \ and\ \bibinfo
  {author} {\bibfnamefont {X.}~\bibnamefont {Chen}},\ }\href {\doibase
  10.1103/PhysRevD.103.L091501} {\bibfield  {journal} {\bibinfo  {journal}
  {Phys. Rev. D}\ }\textbf {\bibinfo {volume} {103}},\ \bibinfo {pages}
  {L091501} (\bibinfo {year} {2021})}\BibitemShut {NoStop}%
\bibitem [{\citenamefont {Hatta}\ \emph {et~al.}(2019)\citenamefont {Hatta},
  \citenamefont {Rajan},\ and\ \citenamefont {Yang}}]{Hatta2}%
  \BibitemOpen
  \bibfield  {author} {\bibinfo {author} {\bibfnamefont {Y.}~\bibnamefont
  {Hatta}}, \bibinfo {author} {\bibfnamefont {A.}~\bibnamefont {Rajan}}, \ and\
  \bibinfo {author} {\bibfnamefont {D.-L.}\ \bibnamefont {Yang}},\ }\href
  {\doibase 10.1103/physrevd.100.014032} {\bibfield  {journal} {\bibinfo
  {journal} {Phys. Rev. D}\ }\textbf {\bibinfo {volume} {100}} (\bibinfo {year}
  {2019}),\ 10.1103/physrevd.100.014032}\BibitemShut {NoStop}%
\bibitem [{\citenamefont {Mamo}\ and\ \citenamefont {Zahed}(2020)}]{Zahed1}%
  \BibitemOpen
  \bibfield  {author} {\bibinfo {author} {\bibfnamefont {K.~A.}\ \bibnamefont
  {Mamo}}\ and\ \bibinfo {author} {\bibfnamefont {I.}~\bibnamefont {Zahed}},\
  }\href {\doibase 10.1103/physrevd.101.086003} {\bibfield  {journal} {\bibinfo
   {journal} {Phys. Rev. D}\ }\textbf {\bibinfo {volume} {101}} (\bibinfo
  {year} {2020}),\ 10.1103/physrevd.101.086003}\BibitemShut {NoStop}%
\bibitem [{\citenamefont {Mamo}\ and\ \citenamefont
  {Zahed}(2021{\natexlab{b}})}]{Zahed3}%
  \BibitemOpen
  \bibfield  {author} {\bibinfo {author} {\bibfnamefont {K.~A.}\ \bibnamefont
  {Mamo}}\ and\ \bibinfo {author} {\bibfnamefont {I.}~\bibnamefont {Zahed}},\
  }\href {\doibase 10.1103/physrevd.104.066023} {\bibfield  {journal} {\bibinfo
   {journal} {Physical Review D}\ }\textbf {\bibinfo {volume} {104}} (\bibinfo
  {year} {2021}{\natexlab{b}}),\ 10.1103/physrevd.104.066023}\BibitemShut
  {NoStop}%
\bibitem [{\citenamefont {Sun}\ \emph {et~al.}(2022)\citenamefont {Sun},
  \citenamefont {Tong},\ and\ \citenamefont {Yuan}}]{Feng2}%
  \BibitemOpen
  \bibfield  {author} {\bibinfo {author} {\bibfnamefont {P.}~\bibnamefont
  {Sun}}, \bibinfo {author} {\bibfnamefont {X.-B.}\ \bibnamefont {Tong}}, \
  and\ \bibinfo {author} {\bibfnamefont {F.}~\bibnamefont {Yuan}},\ }\href
  {\doibase 10.1103/physrevd.105.054032} {\bibfield  {journal} {\bibinfo
  {journal} {Phys. Rev. D}\ }\textbf {\bibinfo {volume} {105}} (\bibinfo {year}
  {2022}),\ 10.1103/physrevd.105.054032}\BibitemShut {NoStop}%
\bibitem [{\citenamefont {Du}\ \emph {et~al.}(2020)\citenamefont {Du},
  \citenamefont {Baru}, \citenamefont {Guo}, \citenamefont {Hanhart},
  \citenamefont {Meißner}, \citenamefont {Nefediev},\ and\ \citenamefont
  {Strakovsky}}]{openc}%
  \BibitemOpen
  \bibfield  {author} {\bibinfo {author} {\bibfnamefont {M.-L.}\ \bibnamefont
  {Du}}, \bibinfo {author} {\bibfnamefont {V.}~\bibnamefont {Baru}}, \bibinfo
  {author} {\bibfnamefont {F.-K.}\ \bibnamefont {Guo}}, \bibinfo {author}
  {\bibfnamefont {C.}~\bibnamefont {Hanhart}}, \bibinfo {author} {\bibfnamefont
  {U.-G.}\ \bibnamefont {Meißner}}, \bibinfo {author} {\bibfnamefont
  {A.}~\bibnamefont {Nefediev}}, \ and\ \bibinfo {author} {\bibfnamefont
  {I.}~\bibnamefont {Strakovsky}},\ }\href {\doibase
  10.1140/epjc/s10052-020-08620-5} {\bibfield  {journal} {\bibinfo  {journal}
  {Eur.Phys.J.C}\ }\textbf {\bibinfo {volume} {80}} (\bibinfo {year} {2020}),\
  10.1140/epjc/s10052-020-08620-5}\BibitemShut {NoStop}%
\bibitem [{\citenamefont {Aaij}\ \emph {et~al.}(2015)\citenamefont {Aaij} \emph
  {et~al.}}]{LHCb1}%
  \BibitemOpen
  \bibfield  {author} {\bibinfo {author} {\bibfnamefont {R.}~\bibnamefont
  {Aaij}} \emph {et~al.} (\bibinfo {collaboration} {LHCb Collaboration}),\
  }\href@noop {} {\bibfield  {journal} {\bibinfo  {journal} {Phys. Rev. Lett.}\
  }\textbf {\bibinfo {volume} {115}},\ \bibinfo {pages} {072001} (\bibinfo
  {year} {2015})}\BibitemShut {NoStop}%
\bibitem [{\citenamefont {Aaij}\ \emph {et~al.}(2019)\citenamefont {Aaij} \emph
  {et~al.}}]{LHCb3}%
  \BibitemOpen
  \bibfield  {author} {\bibinfo {author} {\bibfnamefont {R.}~\bibnamefont
  {Aaij}} \emph {et~al.} (\bibinfo {collaboration} {LHCb Collaboration}),\
  }\href {\doibase 10.1103/PhysRevLett.122.222001} {\bibfield  {journal}
  {\bibinfo  {journal} {Phys. Rev. Lett.}\ }\textbf {\bibinfo {volume} {122}},\
  \bibinfo {pages} {222001} (\bibinfo {year} {2019})}\BibitemShut {NoStop}%
\bibitem [{\citenamefont {Wang}\ \emph {et~al.}(2015)\citenamefont {Wang},
  \citenamefont {Liu},\ and\ \citenamefont {Zhao}}]{Wang}%
  \BibitemOpen
  \bibfield  {author} {\bibinfo {author} {\bibfnamefont {Q.}~\bibnamefont
  {Wang}}, \bibinfo {author} {\bibfnamefont {X.-H.}\ \bibnamefont {Liu}}, \
  and\ \bibinfo {author} {\bibfnamefont {Q.}~\bibnamefont {Zhao}},\ }\href
  {\doibase 10.1103/PhysRevD.92.034022} {\bibfield  {journal} {\bibinfo
  {journal} {Phys. Rev. D}\ }\textbf {\bibinfo {volume} {92}},\ \bibinfo
  {pages} {034022} (\bibinfo {year} {2015})}\BibitemShut {NoStop}%
\bibitem [{\citenamefont {Kubarovsky}\ and\ \citenamefont
  {Voloshin}(2015)}]{Kubarovsky}%
  \BibitemOpen
  \bibfield  {author} {\bibinfo {author} {\bibfnamefont {V.}~\bibnamefont
  {Kubarovsky}}\ and\ \bibinfo {author} {\bibfnamefont {M.~B.}\ \bibnamefont
  {Voloshin}},\ }\href@noop {} {\bibfield  {journal} {\bibinfo  {journal}
  {Phys. Rev. D}\ }\textbf {\bibinfo {volume} {92}},\ \bibinfo {pages} {031502}
  (\bibinfo {year} {2015})}\BibitemShut {NoStop}%
\bibitem [{\citenamefont {Karliner}\ and\ \citenamefont
  {Rosner}(2016)}]{Karliner}%
  \BibitemOpen
  \bibfield  {author} {\bibinfo {author} {\bibfnamefont {M.}~\bibnamefont
  {Karliner}}\ and\ \bibinfo {author} {\bibfnamefont {J.}~\bibnamefont
  {Rosner}},\ }\href@noop {} {\bibfield  {journal} {\bibinfo  {journal} {Phys.
  Lett. B}\ }\textbf {\bibinfo {volume} {752}},\ \bibinfo {pages} {329}
  (\bibinfo {year} {2016})}\BibitemShut {NoStop}%
\bibitem [{\citenamefont {Blin}\ \emph {et~al.}(2016)\citenamefont {Blin},
  \citenamefont {{Fernandez - Ramirez}}, \citenamefont {Jackura}, \citenamefont
  {Mathieu}, \citenamefont {Mokeev}, \citenamefont {Pilloni},\ and\
  \citenamefont {Szczepaniak}}]{Blin}%
  \BibitemOpen
  \bibfield  {author} {\bibinfo {author} {\bibfnamefont {A.}~\bibnamefont
  {Blin}}, \bibinfo {author} {\bibfnamefont {C.}~\bibnamefont {{Fernandez -
  Ramirez}}}, \bibinfo {author} {\bibfnamefont {A.}~\bibnamefont {Jackura}},
  \bibinfo {author} {\bibfnamefont {V.}~\bibnamefont {Mathieu}}, \bibinfo
  {author} {\bibfnamefont {V.}~\bibnamefont {Mokeev}}, \bibinfo {author}
  {\bibfnamefont {A.}~\bibnamefont {Pilloni}}, \ and\ \bibinfo {author}
  {\bibfnamefont {A.}~\bibnamefont {Szczepaniak}},\ }\href@noop {} {\bibfield
  {journal} {\bibinfo  {journal} {Phys. Rev. D}\ }\textbf {\bibinfo {volume}
  {94}},\ \bibinfo {pages} {034002} (\bibinfo {year} {2016})}\BibitemShut
  {NoStop}%
\bibitem [{\citenamefont {Adhikari}\ \emph {et~al.}(2021)\citenamefont
  {Adhikari}, \citenamefont {Akondi}, \citenamefont {Al~Ghoul}, \citenamefont
  {Ali}, \citenamefont {Amaryan}, \citenamefont {Anassontzis}, \citenamefont
  {Austregesilo}, \citenamefont {Barbosa}, \citenamefont {Barlow},
  \citenamefont {Barnes},\ and\ \citenamefont {et~al.}}]{GlueX_NIM}%
  \BibitemOpen
  \bibfield  {author} {\bibinfo {author} {\bibfnamefont {S.}~\bibnamefont
  {Adhikari}}, \bibinfo {author} {\bibfnamefont {C.}~\bibnamefont {Akondi}},
  \bibinfo {author} {\bibfnamefont {H.}~\bibnamefont {Al~Ghoul}}, \bibinfo
  {author} {\bibfnamefont {A.}~\bibnamefont {Ali}}, \bibinfo {author}
  {\bibfnamefont {M.}~\bibnamefont {Amaryan}}, \bibinfo {author} {\bibfnamefont
  {E.}~\bibnamefont {Anassontzis}}, \bibinfo {author} {\bibfnamefont
  {A.}~\bibnamefont {Austregesilo}}, \bibinfo {author} {\bibfnamefont
  {F.}~\bibnamefont {Barbosa}}, \bibinfo {author} {\bibfnamefont
  {J.}~\bibnamefont {Barlow}}, \bibinfo {author} {\bibfnamefont
  {A.}~\bibnamefont {Barnes}}, \ and\ \bibinfo {author} {\bibnamefont
  {et~al.}},\ }\href {\doibase 10.1016/j.nima.2020.164807} {\bibfield
  {journal} {\bibinfo  {journal} {Nucl. Instrum. Meth. A}\ }\textbf {\bibinfo
  {volume} {987}},\ \bibinfo {pages} {164807} (\bibinfo {year}
  {2021})}\BibitemShut {NoStop}%
\bibitem [{\citenamefont {Barbosa}\ \emph {et~al.}(2015)\citenamefont
  {Barbosa}, \citenamefont {Hutton}, \citenamefont {Sitnikov}, \citenamefont
  {Somov}, \citenamefont {Somov},\ and\ \citenamefont {Tolstukhin}}]{PS}%
  \BibitemOpen
  \bibfield  {author} {\bibinfo {author} {\bibfnamefont {F.}~\bibnamefont
  {Barbosa}}, \bibinfo {author} {\bibfnamefont {C.}~\bibnamefont {Hutton}},
  \bibinfo {author} {\bibfnamefont {A.}~\bibnamefont {Sitnikov}}, \bibinfo
  {author} {\bibfnamefont {A.}~\bibnamefont {Somov}}, \bibinfo {author}
  {\bibfnamefont {S.}~\bibnamefont {Somov}}, \ and\ \bibinfo {author}
  {\bibfnamefont {I.}~\bibnamefont {Tolstukhin}},\ }\href@noop {} {\bibfield
  {journal} {\bibinfo  {journal} {Nucl. Instrum. Meth. A}\ }\textbf {\bibinfo
  {volume} {795}},\ \bibinfo {pages} {376} (\bibinfo {year}
  {2015})}\BibitemShut {NoStop}%
\bibitem [{\citenamefont {Pooser}\ \emph {et~al.}(2019)\citenamefont {Pooser},
  \citenamefont {Barbosa}, \citenamefont {Boeglin}, \citenamefont {Hutton},
  \citenamefont {Ito}, \citenamefont {Kamel}, \citenamefont {LLodra},
  \citenamefont {Sandoval}, \citenamefont {Taylor}, \citenamefont {Whitlatch},
  \citenamefont {Worthington}, \citenamefont {Yero},\ and\ \citenamefont
  {Zihlmann}}]{ST}%
  \BibitemOpen
  \bibfield  {author} {\bibinfo {author} {\bibfnamefont {E.}~\bibnamefont
  {Pooser}}, \bibinfo {author} {\bibfnamefont {F.}~\bibnamefont {Barbosa}},
  \bibinfo {author} {\bibfnamefont {W.}~\bibnamefont {Boeglin}}, \bibinfo
  {author} {\bibfnamefont {C.}~\bibnamefont {Hutton}}, \bibinfo {author}
  {\bibfnamefont {M.}~\bibnamefont {Ito}}, \bibinfo {author} {\bibfnamefont
  {M.}~\bibnamefont {Kamel}}, \bibinfo {author} {\bibfnamefont {P.~K.~A.}\
  \bibnamefont {LLodra}}, \bibinfo {author} {\bibfnamefont {N.}~\bibnamefont
  {Sandoval}}, \bibinfo {author} {\bibfnamefont {S.}~\bibnamefont {Taylor}},
  \bibinfo {author} {\bibfnamefont {T.}~\bibnamefont {Whitlatch}}, \bibinfo
  {author} {\bibfnamefont {S.}~\bibnamefont {Worthington}}, \bibinfo {author}
  {\bibfnamefont {C.}~\bibnamefont {Yero}}, \ and\ \bibinfo {author}
  {\bibfnamefont {B.}~\bibnamefont {Zihlmann}},\ }\href@noop {} {\bibfield
  {journal} {\bibinfo  {journal} {Nucl. Instrum. Meth. A}\ }\textbf {\bibinfo
  {volume} {927}},\ \bibinfo {pages} {330} (\bibinfo {year}
  {2019})}\BibitemShut {NoStop}%
\bibitem [{\citenamefont {Jarvis}\ \emph {et~al.}(2020)\citenamefont {Jarvis},
  \citenamefont {Meyer}, \citenamefont {Zihlmann}, \citenamefont {Staib},
  \citenamefont {Austregesilo}, \citenamefont {Barbosa}, \citenamefont
  {Dickover}, \citenamefont {Razmyslovich}, \citenamefont {Taylor},
  \citenamefont {Van~Haarlem} \emph {et~al.}}]{CDC}%
  \BibitemOpen
  \bibfield  {author} {\bibinfo {author} {\bibfnamefont {N.}~\bibnamefont
  {Jarvis}}, \bibinfo {author} {\bibfnamefont {C.}~\bibnamefont {Meyer}},
  \bibinfo {author} {\bibfnamefont {B.}~\bibnamefont {Zihlmann}}, \bibinfo
  {author} {\bibfnamefont {M.}~\bibnamefont {Staib}}, \bibinfo {author}
  {\bibfnamefont {A.}~\bibnamefont {Austregesilo}}, \bibinfo {author}
  {\bibfnamefont {F.}~\bibnamefont {Barbosa}}, \bibinfo {author} {\bibfnamefont
  {C.}~\bibnamefont {Dickover}}, \bibinfo {author} {\bibfnamefont
  {V.}~\bibnamefont {Razmyslovich}}, \bibinfo {author} {\bibfnamefont
  {S.}~\bibnamefont {Taylor}}, \bibinfo {author} {\bibfnamefont
  {Y.}~\bibnamefont {Van~Haarlem}},  \emph {et~al.},\ }\href {\doibase
  10.1016/j.nima.2020.163727} {\bibfield  {journal} {\bibinfo  {journal} {Nucl.
  Instrum. Meth. A}\ }\textbf {\bibinfo {volume} {962}},\ \bibinfo {pages}
  {163727} (\bibinfo {year} {2020})}\BibitemShut {NoStop}%
\bibitem [{\citenamefont {Pentchev}\ \emph {et~al.}(2017)\citenamefont
  {Pentchev}, \citenamefont {Barbosa}, \citenamefont {Berdnikov}, \citenamefont
  {Butler}, \citenamefont {Furletov}, \citenamefont {Robison},\ and\
  \citenamefont {Zihlmann}}]{FDC}%
  \BibitemOpen
  \bibfield  {author} {\bibinfo {author} {\bibfnamefont {L.}~\bibnamefont
  {Pentchev}}, \bibinfo {author} {\bibfnamefont {F.}~\bibnamefont {Barbosa}},
  \bibinfo {author} {\bibfnamefont {V.}~\bibnamefont {Berdnikov}}, \bibinfo
  {author} {\bibfnamefont {D.}~\bibnamefont {Butler}}, \bibinfo {author}
  {\bibfnamefont {S.}~\bibnamefont {Furletov}}, \bibinfo {author}
  {\bibfnamefont {L.}~\bibnamefont {Robison}}, \ and\ \bibinfo {author}
  {\bibfnamefont {B.}~\bibnamefont {Zihlmann}},\ }\href@noop {} {\bibfield
  {journal} {\bibinfo  {journal} {Nucl. Instrum. Meth. A}\ }\textbf {\bibinfo
  {volume} {845}},\ \bibinfo {pages} {281} (\bibinfo {year}
  {2017})}\BibitemShut {NoStop}%
\bibitem [{\citenamefont {Beattie}\ \emph {et~al.}(2018)\citenamefont
  {Beattie}, \citenamefont {Foda}, \citenamefont {Henschel}, \citenamefont
  {Katsaganis}, \citenamefont {Krueger}, \citenamefont {Lolos}, \citenamefont
  {Papandreou} \emph {et~al.}}]{BCAL}%
  \BibitemOpen
  \bibfield  {author} {\bibinfo {author} {\bibfnamefont {T.~D.}\ \bibnamefont
  {Beattie}}, \bibinfo {author} {\bibfnamefont {A.~M.}\ \bibnamefont {Foda}},
  \bibinfo {author} {\bibfnamefont {C.~L.}\ \bibnamefont {Henschel}}, \bibinfo
  {author} {\bibfnamefont {S.}~\bibnamefont {Katsaganis}}, \bibinfo {author}
  {\bibfnamefont {S.~T.}\ \bibnamefont {Krueger}}, \bibinfo {author}
  {\bibfnamefont {G.~J.}\ \bibnamefont {Lolos}}, \bibinfo {author}
  {\bibfnamefont {Z.}~\bibnamefont {Papandreou}},  \emph {et~al.},\ }\href@noop
  {} {\bibfield  {journal} {\bibinfo  {journal} {Nucl. Instrum. Meth. A}\
  }\textbf {\bibinfo {volume} {896}},\ \bibinfo {pages} {24} (\bibinfo {year}
  {2018})}\BibitemShut {NoStop}%
\bibitem [{\citenamefont {Paremuzyan}(2017)}]{Rafo}%
  \BibitemOpen
  \bibfield  {author} {\bibinfo {author} {\bibfnamefont {R.}~\bibnamefont
  {Paremuzyan}},\ }\href@noop {} {\bibfield  {journal} {\bibinfo  {journal}
  {(private communication)}\ } (\bibinfo {year} {2017})}\BibitemShut {NoStop}%
\bibitem [{\citenamefont {Berger}\ \emph {et~al.}(2002)\citenamefont {Berger},
  \citenamefont {Diehl},\ and\ \citenamefont {Pire}}]{Berger}%
  \BibitemOpen
  \bibfield  {author} {\bibinfo {author} {\bibfnamefont {E.}~\bibnamefont
  {Berger}}, \bibinfo {author} {\bibfnamefont {M.}~\bibnamefont {Diehl}}, \
  and\ \bibinfo {author} {\bibfnamefont {B.}~\bibnamefont {Pire}},\ }\href@noop
  {} {\bibfield  {journal} {\bibinfo  {journal} {Eur.Phys.J.C}\ }\textbf
  {\bibinfo {volume} {23}},\ \bibinfo {pages} {675} (\bibinfo {year}
  {2002})}\BibitemShut {NoStop}%
\bibitem [{\citenamefont {Borah}\ \emph {et~al.}(2020)\citenamefont {Borah},
  \citenamefont {Hill}, \citenamefont {Lee},\ and\ \citenamefont
  {Tomalak}}]{protonFF}%
  \BibitemOpen
  \bibfield  {author} {\bibinfo {author} {\bibfnamefont {K.}~\bibnamefont
  {Borah}}, \bibinfo {author} {\bibfnamefont {R.~J.}\ \bibnamefont {Hill}},
  \bibinfo {author} {\bibfnamefont {G.}~\bibnamefont {Lee}}, \ and\ \bibinfo
  {author} {\bibfnamefont {O.}~\bibnamefont {Tomalak}},\ }\href {\doibase
  10.1103/physrevd.102.074012} {\bibfield  {journal} {\bibinfo  {journal}
  {Phys. Rev. D}\ }\textbf {\bibinfo {volume} {102}} (\bibinfo {year} {2020}),\
  10.1103/physrevd.102.074012}\BibitemShut {NoStop}%
\bibitem [{\citenamefont {Allison}\ \emph {et~al.}(2016)\citenamefont {Allison}
  \emph {et~al.}}]{GEANT4}%
  \BibitemOpen
  \bibfield  {author} {\bibinfo {author} {\bibfnamefont {J.}~\bibnamefont
  {Allison}} \emph {et~al.},\ }\href {\doibase 10.1016/j.nima.2016.06.125}
  {\bibfield  {journal} {\bibinfo  {journal} {Nucl. Instrum. Meth. A}\ }\textbf
  {\bibinfo {volume} {835}},\ \bibinfo {pages} {186} (\bibinfo {year}
  {2016})}\BibitemShut {NoStop}%
\bibitem [{\citenamefont {Verkerke}\ and\ \citenamefont {Kirkby}()}]{RooFit}%
  \BibitemOpen
  \bibfield  {author} {\bibinfo {author} {\bibfnamefont {W.}~\bibnamefont
  {Verkerke}}\ and\ \bibinfo {author} {\bibfnamefont {D.}~\bibnamefont
  {Kirkby}},\ }\href@noop {} {}\bibinfo {howpublished}
  {\url{https://root.cern/download/doc/RooFit_Users_Manual_2.91-33.pdf}}\BibitemShut
  {NoStop}%
\bibitem [{\citenamefont {Tanabashi~{et al. (Particle Data
  Group)}}(2018)}]{pdg}%
  \BibitemOpen
  \bibfield  {author} {\bibinfo {author} {\bibfnamefont {M.}~\bibnamefont
  {Tanabashi~{et al. (Particle Data Group)}}},\ }\href@noop {} {\bibfield
  {journal} {\bibinfo  {journal} {Phys. Rev. D}\ }\textbf {\bibinfo {volume}
  {98}},\ \bibinfo {pages} {030001} (\bibinfo {year} {2018})}\BibitemShut
  {NoStop}%
\bibitem [{\citenamefont {Heller}\ \emph {et~al.}(2018)\citenamefont {Heller},
  \citenamefont {Tomalak},\ and\ \citenamefont
  {Vanderhaeghen}}]{Vanderhaeghen}%
  \BibitemOpen
  \bibfield  {author} {\bibinfo {author} {\bibfnamefont {M.}~\bibnamefont
  {Heller}}, \bibinfo {author} {\bibfnamefont {O.}~\bibnamefont {Tomalak}}, \
  and\ \bibinfo {author} {\bibfnamefont {M.}~\bibnamefont {Vanderhaeghen}},\
  }\href@noop {} {\bibfield  {journal} {\bibinfo  {journal} {Phys. Rev. D}\
  }\textbf {\bibinfo {volume} {97}},\ \bibinfo {pages} {076012} (\bibinfo
  {year} {2018})}\BibitemShut {NoStop}%
\bibitem [{\citenamefont {Boer}(2019)}]{Marie_note}%
  \BibitemOpen
  \bibfield  {author} {\bibinfo {author} {\bibfnamefont {M.}~\bibnamefont
  {Boer}},\ }\href@noop {} {\bibfield  {journal} {\bibinfo  {journal}
  {Jefferson Lab, Hall C public note no. 1000}\ } (\bibinfo {year}
  {2019})}\BibitemShut {NoStop}%
\bibitem [{\citenamefont {Boer}\ \emph {et~al.}(2015)\citenamefont {Boer},
  \citenamefont {Guidal},\ and\ \citenamefont {Vanderhaeghen}}]{Marie_journal}%
  \BibitemOpen
  \bibfield  {author} {\bibinfo {author} {\bibfnamefont {M.}~\bibnamefont
  {Boer}}, \bibinfo {author} {\bibfnamefont {M.}~\bibnamefont {Guidal}}, \ and\
  \bibinfo {author} {\bibfnamefont {M.}~\bibnamefont {Vanderhaeghen}},\
  }\href@noop {} {\bibfield  {journal} {\bibinfo  {journal} {Eur.Phys.J.A}\
  }\textbf {\bibinfo {volume} {51}},\ \bibinfo {pages} {103} (\bibinfo {year}
  {2015})}\BibitemShut {NoStop}%
\bibitem [{\citenamefont {Duran}\ \emph {et~al.}(2023)\citenamefont {Duran},
  \citenamefont {Meziani}, \citenamefont {Joosten}, \citenamefont {Jones},
  \citenamefont {Prasad}, \citenamefont {Peng}, \citenamefont {Armstrong},
  \citenamefont {Atac}, \citenamefont {Chudakov}, \citenamefont {Bhatt},
  \citenamefont {Bhetuwal}, \citenamefont {Boer}, \citenamefont {Camsonne},
  \citenamefont {Chen}, \citenamefont {Dalton}, \citenamefont {Deokar},
  \citenamefont {Diefenthaler}, \citenamefont {Dunne}, \citenamefont
  {El~Fassi}, \citenamefont {Fuchey}, \citenamefont {Gao}, \citenamefont
  {Gaskell}, \citenamefont {Hansen}, \citenamefont {Hauenstein}, \citenamefont
  {Higinbotham}, \citenamefont {Jia}, \citenamefont {Karki}, \citenamefont
  {Keppel}, \citenamefont {King}, \citenamefont {Ko}, \citenamefont {Li},
  \citenamefont {Li}, \citenamefont {Mack}, \citenamefont {Malace},
  \citenamefont {McCaughan}, \citenamefont {McClellan}, \citenamefont
  {Michaels}, \citenamefont {Meekins}, \citenamefont {Paolone}, \citenamefont
  {Pentchev}, \citenamefont {Pooser}, \citenamefont {Puckett}, \citenamefont
  {Radloff}, \citenamefont {Rehfuss}, \citenamefont {Reimer}, \citenamefont
  {Riordan}, \citenamefont {Sawatzky}, \citenamefont {Smith}, \citenamefont
  {Sparveris}, \citenamefont {Szumila-Vance}, \citenamefont {Wood},
  \citenamefont {Xie}, \citenamefont {Ye}, \citenamefont {Yero},\ and\
  \citenamefont {Zhao}}]{hallc_jp007}%
  \BibitemOpen
  \bibfield  {author} {\bibinfo {author} {\bibfnamefont {B.}~\bibnamefont
  {Duran}}, \bibinfo {author} {\bibfnamefont {Z.-E.}\ \bibnamefont {Meziani}},
  \bibinfo {author} {\bibfnamefont {S.}~\bibnamefont {Joosten}}, \bibinfo
  {author} {\bibfnamefont {M.~K.}\ \bibnamefont {Jones}}, \bibinfo {author}
  {\bibfnamefont {S.}~\bibnamefont {Prasad}}, \bibinfo {author} {\bibfnamefont
  {C.}~\bibnamefont {Peng}}, \bibinfo {author} {\bibfnamefont {W.}~\bibnamefont
  {Armstrong}}, \bibinfo {author} {\bibfnamefont {H.}~\bibnamefont {Atac}},
  \bibinfo {author} {\bibfnamefont {E.}~\bibnamefont {Chudakov}}, \bibinfo
  {author} {\bibfnamefont {H.}~\bibnamefont {Bhatt}}, \bibinfo {author}
  {\bibfnamefont {D.}~\bibnamefont {Bhetuwal}}, \bibinfo {author}
  {\bibfnamefont {M.}~\bibnamefont {Boer}}, \bibinfo {author} {\bibfnamefont
  {A.}~\bibnamefont {Camsonne}}, \bibinfo {author} {\bibfnamefont {J.-P.}\
  \bibnamefont {Chen}}, \bibinfo {author} {\bibfnamefont {M.~M.}\ \bibnamefont
  {Dalton}}, \bibinfo {author} {\bibfnamefont {N.}~\bibnamefont {Deokar}},
  \bibinfo {author} {\bibfnamefont {M.}~\bibnamefont {Diefenthaler}}, \bibinfo
  {author} {\bibfnamefont {J.}~\bibnamefont {Dunne}}, \bibinfo {author}
  {\bibfnamefont {L.}~\bibnamefont {El~Fassi}}, \bibinfo {author}
  {\bibfnamefont {E.}~\bibnamefont {Fuchey}}, \bibinfo {author} {\bibfnamefont
  {H.}~\bibnamefont {Gao}}, \bibinfo {author} {\bibfnamefont {D.}~\bibnamefont
  {Gaskell}}, \bibinfo {author} {\bibfnamefont {O.}~\bibnamefont {Hansen}},
  \bibinfo {author} {\bibfnamefont {F.}~\bibnamefont {Hauenstein}}, \bibinfo
  {author} {\bibfnamefont {D.}~\bibnamefont {Higinbotham}}, \bibinfo {author}
  {\bibfnamefont {S.}~\bibnamefont {Jia}}, \bibinfo {author} {\bibfnamefont
  {A.}~\bibnamefont {Karki}}, \bibinfo {author} {\bibfnamefont
  {C.}~\bibnamefont {Keppel}}, \bibinfo {author} {\bibfnamefont
  {P.}~\bibnamefont {King}}, \bibinfo {author} {\bibfnamefont {H.~S.}\
  \bibnamefont {Ko}}, \bibinfo {author} {\bibfnamefont {X.}~\bibnamefont {Li}},
  \bibinfo {author} {\bibfnamefont {R.}~\bibnamefont {Li}}, \bibinfo {author}
  {\bibfnamefont {D.}~\bibnamefont {Mack}}, \bibinfo {author} {\bibfnamefont
  {S.}~\bibnamefont {Malace}}, \bibinfo {author} {\bibfnamefont
  {M.}~\bibnamefont {McCaughan}}, \bibinfo {author} {\bibfnamefont {R.~E.}\
  \bibnamefont {McClellan}}, \bibinfo {author} {\bibfnamefont {R.}~\bibnamefont
  {Michaels}}, \bibinfo {author} {\bibfnamefont {D.}~\bibnamefont {Meekins}},
  \bibinfo {author} {\bibfnamefont {M.}~\bibnamefont {Paolone}}, \bibinfo
  {author} {\bibfnamefont {L.}~\bibnamefont {Pentchev}}, \bibinfo {author}
  {\bibfnamefont {E.}~\bibnamefont {Pooser}}, \bibinfo {author} {\bibfnamefont
  {A.}~\bibnamefont {Puckett}}, \bibinfo {author} {\bibfnamefont
  {R.}~\bibnamefont {Radloff}}, \bibinfo {author} {\bibfnamefont
  {M.}~\bibnamefont {Rehfuss}}, \bibinfo {author} {\bibfnamefont {P.~E.}\
  \bibnamefont {Reimer}}, \bibinfo {author} {\bibfnamefont {S.}~\bibnamefont
  {Riordan}}, \bibinfo {author} {\bibfnamefont {B.}~\bibnamefont {Sawatzky}},
  \bibinfo {author} {\bibfnamefont {A.}~\bibnamefont {Smith}}, \bibinfo
  {author} {\bibfnamefont {N.}~\bibnamefont {Sparveris}}, \bibinfo {author}
  {\bibfnamefont {H.}~\bibnamefont {Szumila-Vance}}, \bibinfo {author}
  {\bibfnamefont {S.}~\bibnamefont {Wood}}, \bibinfo {author} {\bibfnamefont
  {J.}~\bibnamefont {Xie}}, \bibinfo {author} {\bibfnamefont {Z.}~\bibnamefont
  {Ye}}, \bibinfo {author} {\bibfnamefont {C.}~\bibnamefont {Yero}}, \ and\
  \bibinfo {author} {\bibfnamefont {Z.}~\bibnamefont {Zhao}},\ }\href {\doibase
  10.1038/s41586-023-05730-4} {\bibfield  {journal} {\bibinfo  {journal}
  {Nature}\ }\textbf {\bibinfo {volume} {615}},\ \bibinfo {pages} {813}
  (\bibinfo {year} {2023})}\BibitemShut {NoStop}%
\bibitem [{\citenamefont {Gryniuk}\ and\ \citenamefont
  {Vanderhaeghen}(2016)}]{Vanderhaeghen16}%
  \BibitemOpen
  \bibfield  {author} {\bibinfo {author} {\bibfnamefont {O.}~\bibnamefont
  {Gryniuk}}\ and\ \bibinfo {author} {\bibfnamefont {M.}~\bibnamefont
  {Vanderhaeghen}},\ }\href {\doibase 10.1103/PhysRevD.94.074001} {\bibfield
  {journal} {\bibinfo  {journal} {Phys. Rev. D}\ }\textbf {\bibinfo {volume}
  {94}},\ \bibinfo {pages} {074001} (\bibinfo {year} {2016})}\BibitemShut
  {NoStop}%
\bibitem [{\citenamefont {Tong}\ \emph {et~al.}(2021)\citenamefont {Tong},
  \citenamefont {Ma},\ and\ \citenamefont {Yuan}}]{Feng1}%
  \BibitemOpen
  \bibfield  {author} {\bibinfo {author} {\bibfnamefont {X.-B.}\ \bibnamefont
  {Tong}}, \bibinfo {author} {\bibfnamefont {J.-P.}\ \bibnamefont {Ma}}, \ and\
  \bibinfo {author} {\bibfnamefont {F.}~\bibnamefont {Yuan}},\ }\href@noop {}
  {\enquote {\bibinfo {title} {Gluon gravitational form factors at large
  momentum transfer},}\ } (\bibinfo {year} {2021}),\ \Eprint
  {http://arxiv.org/abs/2101.02395} {arXiv:2101.02395 [hep-ph]} \BibitemShut
  {NoStop}%
\bibitem [{\citenamefont {Shanahan}\ and\ \citenamefont
  {Detmold}(2019)}]{Shanahan}%
  \BibitemOpen
  \bibfield  {author} {\bibinfo {author} {\bibfnamefont {P.~E.}\ \bibnamefont
  {Shanahan}}\ and\ \bibinfo {author} {\bibfnamefont {W.}~\bibnamefont
  {Detmold}},\ }\href {\doibase 10.1103/PhysRevD.99.014511} {\bibfield
  {journal} {\bibinfo  {journal} {Phys. Rev. D}\ }\textbf {\bibinfo {volume}
  {99}},\ \bibinfo {pages} {014511} (\bibinfo {year} {2019})},\ \Eprint
  {http://arxiv.org/abs/1810.04626} {arXiv:1810.04626 [hep-lat]} \BibitemShut
  {NoStop}%
\bibitem [{\citenamefont {Camerini}\ \emph {et~al.}(1975)\citenamefont
  {Camerini}, \citenamefont {Learned}, \citenamefont {Prepost}, \citenamefont
  {Spencer}, \citenamefont {Wiser}, \citenamefont {Ash}, \citenamefont
  {Anderson}, \citenamefont {Ritson}, \citenamefont {Sherden},\ and\
  \citenamefont {Sinclair}}]{SLAC}%
  \BibitemOpen
  \bibfield  {author} {\bibinfo {author} {\bibfnamefont {U.}~\bibnamefont
  {Camerini}}, \bibinfo {author} {\bibfnamefont {J.}~\bibnamefont {Learned}},
  \bibinfo {author} {\bibfnamefont {R.}~\bibnamefont {Prepost}}, \bibinfo
  {author} {\bibfnamefont {C.}~\bibnamefont {Spencer}}, \bibinfo {author}
  {\bibfnamefont {D.}~\bibnamefont {Wiser}}, \bibinfo {author} {\bibfnamefont
  {W.}~\bibnamefont {Ash}}, \bibinfo {author} {\bibfnamefont {R.~L.}\
  \bibnamefont {Anderson}}, \bibinfo {author} {\bibfnamefont {D.~M.}\
  \bibnamefont {Ritson}}, \bibinfo {author} {\bibfnamefont {D.}~\bibnamefont
  {Sherden}}, \ and\ \bibinfo {author} {\bibfnamefont {C.~K.}\ \bibnamefont
  {Sinclair}},\ }\href@noop {} {\bibfield  {journal} {\bibinfo  {journal}
  {Phys. Rev. Lett.}\ }\textbf {\bibinfo {volume} {35}},\ \bibinfo {pages}
  {483} (\bibinfo {year} {1975})}\BibitemShut {NoStop}%
\bibitem [{\citenamefont {Pentchev}()}]{QNP2022}%
  \BibitemOpen
  \bibfield  {author} {\bibinfo {author} {\bibfnamefont {L.}~\bibnamefont
  {Pentchev}},\ }\href@noop {} {}\bibinfo {howpublished}
  {\url{https://indico.jlab.org/event/344/contributions/10353/attachments/8401/12067/LPentchev_Jpsi_QNP2022.pdf}}\BibitemShut
  {NoStop}%
\bibitem [{\citenamefont {Gittelman}\ \emph {et~al.}(1975)\citenamefont
  {Gittelman}, \citenamefont {Hanson}, \citenamefont {Larson}, \citenamefont
  {Loh}, \citenamefont {Silverman},\ and\ \citenamefont
  {Theodosiou}}]{Cornell}%
  \BibitemOpen
  \bibfield  {author} {\bibinfo {author} {\bibfnamefont {B.}~\bibnamefont
  {Gittelman}}, \bibinfo {author} {\bibfnamefont {K.~M.}\ \bibnamefont
  {Hanson}}, \bibinfo {author} {\bibfnamefont {D.}~\bibnamefont {Larson}},
  \bibinfo {author} {\bibfnamefont {E.}~\bibnamefont {Loh}}, \bibinfo {author}
  {\bibfnamefont {A.}~\bibnamefont {Silverman}}, \ and\ \bibinfo {author}
  {\bibfnamefont {G.}~\bibnamefont {Theodosiou}},\ }\href@noop {} {\bibfield
  {journal} {\bibinfo  {journal} {Phys. Rev. Lett.}\ }\textbf {\bibinfo
  {volume} {35}},\ \bibinfo {pages} {1616} (\bibinfo {year}
  {1975})}\BibitemShut {NoStop}%
\bibitem [{\citenamefont {Ivanov}\ \emph {et~al.}(2022)\citenamefont {Ivanov},
  \citenamefont {Sznajder}, \citenamefont {Szymanowski},\ and\ \citenamefont
  {Wagner}}]{Lech}%
  \BibitemOpen
  \bibfield  {author} {\bibinfo {author} {\bibfnamefont {D.}~\bibnamefont
  {Ivanov}}, \bibinfo {author} {\bibfnamefont {P.}~\bibnamefont {Sznajder}},
  \bibinfo {author} {\bibfnamefont {L.}~\bibnamefont {Szymanowski}}, \ and\
  \bibinfo {author} {\bibfnamefont {J.}~\bibnamefont {Wagner}},\ }\href@noop {}
  {\bibfield  {journal} {\bibinfo  {journal} {(private communication)}\ }
  (\bibinfo {year} {2022})}\BibitemShut {NoStop}%
\bibitem [{\citenamefont {Sznajder}\ and\ \citenamefont
  {Wagner}(2022)}]{Jakub}%
  \BibitemOpen
  \bibfield  {author} {\bibinfo {author} {\bibfnamefont {P.}~\bibnamefont
  {Sznajder}}\ and\ \bibinfo {author} {\bibfnamefont {J.}~\bibnamefont
  {Wagner}},\ }\href@noop {} {\bibfield  {journal} {\bibinfo  {journal}
  {(private communication)}\ } (\bibinfo {year} {2022})}\BibitemShut {NoStop}%
\bibitem [{\citenamefont {Pire}\ \emph {et~al.}(2022)\citenamefont {Pire},
  \citenamefont {Semenov-Tian-Shansky}, \citenamefont {Shaikhutdinova},\ and\
  \citenamefont {Szymanowski}}]{Lech_u_channel}%
  \BibitemOpen
  \bibfield  {author} {\bibinfo {author} {\bibfnamefont {B.}~\bibnamefont
  {Pire}}, \bibinfo {author} {\bibfnamefont {K.~M.}\ \bibnamefont
  {Semenov-Tian-Shansky}}, \bibinfo {author} {\bibfnamefont {A.~A.}\
  \bibnamefont {Shaikhutdinova}}, \ and\ \bibinfo {author} {\bibfnamefont
  {L.}~\bibnamefont {Szymanowski}},\ }\href {\doibase
  10.48550/ARXIV.2212.07688} {\enquote {\bibinfo {title} {Pion and photon beam
  initiated backward charmonium or lepton pair production},}\ } (\bibinfo
  {year} {2022})\BibitemShut {NoStop}%
\bibitem [{\citenamefont {Pire}\ \emph {et~al.}(2021)\citenamefont {Pire},
  \citenamefont {Semenov-Tian-Shansky},\ and\ \citenamefont
  {Szymanowski}}]{TDA}%
  \BibitemOpen
  \bibfield  {author} {\bibinfo {author} {\bibfnamefont {B.}~\bibnamefont
  {Pire}}, \bibinfo {author} {\bibfnamefont {K.}~\bibnamefont
  {Semenov-Tian-Shansky}}, \ and\ \bibinfo {author} {\bibfnamefont
  {L.}~\bibnamefont {Szymanowski}},\ }\href {\doibase
  10.1016/j.physrep.2021.09.002} {\bibfield  {journal} {\bibinfo  {journal}
  {Phys. Rep.}\ }\textbf {\bibinfo {volume} {940}},\ \bibinfo {pages} {1}
  (\bibinfo {year} {2021})}\BibitemShut {NoStop}%
\bibitem [{\citenamefont {Winney}\ \emph {et~al.}(2023)\citenamefont {Winney},
  \citenamefont {Fernandez-Ramirez}, \citenamefont {Pilloni}, \citenamefont
  {Blin}, \citenamefont {Albaladejo}, \citenamefont {Bibrzycki}, \citenamefont
  {Hammoud}, \citenamefont {Liao}, \citenamefont {Mathieu}, \citenamefont
  {Montana}, \citenamefont {Perry}, \citenamefont {Shastry}, \citenamefont
  {Smith},\ and\ \citenamefont {Szczepaniak}}]{JPAC_paper}%
  \BibitemOpen
  \bibfield  {author} {\bibinfo {author} {\bibfnamefont {D.}~\bibnamefont
  {Winney}}, \bibinfo {author} {\bibfnamefont {C.}~\bibnamefont
  {Fernandez-Ramirez}}, \bibinfo {author} {\bibfnamefont {A.}~\bibnamefont
  {Pilloni}}, \bibinfo {author} {\bibfnamefont {A.~N.~H.}\ \bibnamefont
  {Blin}}, \bibinfo {author} {\bibfnamefont {M.}~\bibnamefont {Albaladejo}},
  \bibinfo {author} {\bibfnamefont {L.}~\bibnamefont {Bibrzycki}}, \bibinfo
  {author} {\bibfnamefont {N.}~\bibnamefont {Hammoud}}, \bibinfo {author}
  {\bibfnamefont {J.}~\bibnamefont {Liao}}, \bibinfo {author} {\bibfnamefont
  {V.}~\bibnamefont {Mathieu}}, \bibinfo {author} {\bibfnamefont
  {G.}~\bibnamefont {Montana}}, \bibinfo {author} {\bibfnamefont {R.~J.}\
  \bibnamefont {Perry}}, \bibinfo {author} {\bibfnamefont {V.}~\bibnamefont
  {Shastry}}, \bibinfo {author} {\bibfnamefont {W.~A.}\ \bibnamefont {Smith}},
  \ and\ \bibinfo {author} {\bibfnamefont {A.~P.}\ \bibnamefont
  {Szczepaniak}},\ }\href@noop {} {\enquote {\bibinfo {title} {Dynamics in
  near-threshold $j/\psi$ photoproduction},}\ } (\bibinfo {year} {2023}),\
  \Eprint {http://arxiv.org/abs/2305.01449} {arXiv:2305.01449 [hep-ph]}
  \BibitemShut {NoStop}%
\end{thebibliography}%
